\documentclass[prb, superscriptaddress, showpacs]{revtex4}
\usepackage{graphicx}
\usepackage{color}
\usepackage{amsmath}
\usepackage{amsfonts}
\usepackage{enumitem}
\usepackage{hyperref}
\hypersetup{
	colorlinks = true,
	citecolor = blue
}
\usepackage{color}

\newcommand{\be}{\begin{equation}}
\newcommand{\ee}{\end{equation}}

\newcommand{\bea}{\begin{eqnarray}}
\newcommand{\eea}{\end{eqnarray}}

\newcommand{\p}{\partial}

\newcommand{\la}{\left\langle}
\newcommand{\ra}{\right\rangle}

\renewcommand{\vec}[1]{{\boldsymbol #1}}

\renewcommand{\phi}{\varphi}
\renewcommand{\epsilon}{\varepsilon}
\def\nn{\nonumber\\}

\usepackage[normalem]{ulem}

\begin{document}

\title{Electronic states of pseudospin-1 fermions in dice lattice ribbons}
\date{\today}

\author{D. O. Oriekhov}
\affiliation{Department of Physics, Taras Shevchenko National University of Kyiv, Kyiv, 03680, Ukraine}

\author{E. V. Gorbar}
\affiliation{Department of Physics, Taras Shevchenko National University of Kyiv, Kyiv, 03680, Ukraine}
\affiliation{Bogolyubov Institute for Theoretical Physics, Kyiv, 03680, Ukraine}

\author{V. P. Gusynin}
\affiliation{Bogolyubov Institute for Theoretical Physics, Kyiv, 03680, Ukraine}

\begin{abstract}

Boundary conditions for the two-dimensional fermions in ribbons of the hexagonal lattice are studied in the dice model whose energy spectrum in
infinite system consists of three bands with one completely flat band of zero energy. Like in graphene the regular lattice terminations
are of the armchair and zigzag types. However, there are four possible zigzag edge terminations in contrast to graphene where only
one type of zigzag termination is possible. Determining the boundary conditions for these lattice terminations, the energy spectra of
pseudospin-1 fermions in dice model ribbons with zigzag and armchair boundary conditions are found. It is shown that the energy levels
for armchair ribbons display the same features as in graphene except the zero energy flat band inherent to the dice model. In addition,
unlike graphene, there are no propagating edge states localized at zigzag boundary and there are specific zigzag terminations which
give rise to bulk modes of a metallic type in dice model ribbons. We find that the existence of the flat zero-energy band in the dice
model is very robust and is not affected by the zigzag and armchair boundaries.
\end{abstract}
\maketitle

\section{Introduction}

After the experimental discovery of graphene [\onlinecite{Novoselov}] there was an explosion of activity in the study of materials
with relativistic like spectrum of quasiparticles whose dynamics is governed by the Dirac or Weyl equation. In addition to graphene, they
are topological insulators [\onlinecite{Kane,Qi}] and 3D Dirac and Weyl semimetals [\onlinecite{Felser,Hasan,Armitage}]. However, the
properties and energy dispersion of the electron states in condensed matter systems are constrained by the crystal space group rather than
the Poincare group. This gives rises to the possibility of fermionic excitations with no analogues in high-energy physics. Indeed, it was
proposed [\onlinecite{Bradlyn}] that the three non-symmorphic space groups host fermionic excitations with three-fold degeneracies. The
corresponding touchings of three bands are topologically non-trivial and either carry a Chern number $\pm 2$ or sit at the critical point
separating the two Chern insulators.

The triply degenerate fermions with nodal points located closely to the Fermi surface were predicted in the
$\text{RERh}_6\text{Ge}_4$ ($\text{RE}=(\text{Y},\text{La},\text{Lu})$) [\onlinecite{Guo}] and $\text{NaCu}_3\text{Te}_2$ [\onlinecite{Xia}]
compounds. They are expected to occur at a high symmetry point in the Brillouin zone and are protected by nonsymmorphic symmetry
[\onlinecite{Bradlyn}]. Latter they were suggested also to exist at a symmetric axis [\onlinecite{Zhu}]-[\onlinecite{Wang}]. Experimentally,
the three-component fermions were observed in MoP and WC [\onlinecite{Lv, Ma}]. The triply degenerate topological semimetals provide an
interesting platform for studying exotic physical properties such as the Fermi arcs, transport anomalies, and topological Lifshitz transitions.
The pairing problem in materials with three bands crossing was studied in Ref.[\onlinecite{Lin}]. A pressure induced superconductivity was
reported in MoP [\onlinecite{Chi}].

Certain lattice systems possess strictly flat bands [\onlinecite{Heikkila}] (for a recent review of artificial flat band systems, see
Ref.[\onlinecite{Flach}]). The dice model provides the historically first example of such a system. It is a tight-binding model of
two-dimensional fermions living on the so-called ${\cal T}_3$ (or dice) lattice where atoms are situated at both the vertices of a hexagonal
lattice and the hexagons centers [\onlinecite{Sutherland,Vidal}]. Since the dice model has three sites per unit cell, the electron
states in this model are described by three-component fermions. It is natural then that the spectrum of the model is comprised of three bands.
The two of them form a Dirac cone and the third band is completely flat and has zero energy [\onlinecite{Raoux}]. All three bands meet at the
$K$ and $K^{\prime}$ points, which are situated at the corners of the Brillouin zone. The ${\cal T}_3$ lattice has been experimentally
realized in Josephson arrays [\onlinecite{Serret}] and its optical realization by laser beams was proposed in Ref.[\onlinecite{Rizzi}].

In linear order to momentum deviations from the $K$ and $K^{\prime}$ points, the low-energy Hamiltonian of the dice model describes
massless pseudospin-1 fermions. These fermions give a surprising strong paramagnetic response in a magnetic field [\onlinecite{Raoux}].
The minimal conductivity and topological Berry winding were analyzed in three-band semimetals in Ref.[\onlinecite{Louvet}].
The dynamic polarizability of the dice model was calculated in the random phase approximation [\onlinecite{Nicol}] and it was found
that the plasmon branch due to strong screening in the flat band is pinched to the point $\omega=|\mathbf{k}|=\mu$. In addition, the
singular nature of the Lindhard function leads to much faster decay of the Friedel oscillations. The magneto-optical conductivity of
pseudospin-1 fermions was calculated in Ref.[\onlinecite{Malcolm}].

Perfectly flat bands are expected to be not stable with respect to generic perturbations. The presence of boundaries is one of such
perturbations. The question whether the flat band survives in finite size systems provides one of the main motivation for the
present study. To answer this question, we consider the two-dimensional dice lattice model, determine the possible types of its
terminations and the corresponding boundary conditions. Then we find the energy spectra and electron states in the dice model
ribbons.

The paper is organized as follows. The dice model and its electron states in infinite system are described in Sec.\ref{sec:model}.
The electron states and energy spectra in ribbons with zigzag and armchair edges are studied in Secs.\ref{sec:zigzag-simple} and
\ref{sec:armchair}, respectively. The results are summarized in Sec.\ref{sec:summary}. The general form of the boundary condition for
the electric current is considered in Appendices \ref{sec:M_parametric} and \ref{appendix:zero-current}.

\section{Model and boundary condition for current}
\label{sec:model}

\begin{figure}
	\includegraphics[scale=0.4]{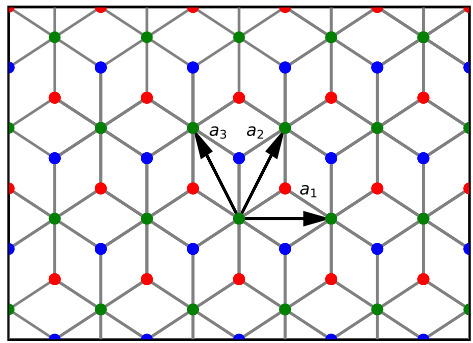}
	\caption{The ${\cal T}_3$ lattice whose red points display the atoms of the
		$A$ sublattice, the blue points describe the $B$ sublattice, and the green points define the $C$ sublattice. The vectors
		$\vec{a}_1=(\sqrt{3},\,0)a$ and $\vec{a}_2=(\sqrt{3}/2,\,3/2)a$ are the basis vectors of the $C$ sublattice.}
	\label{fig1}
\end{figure}

The dice model describes quasiparticles in two dimensions with pseudospin $S=1$ on the $\mathcal{T}_3$ lattice schematically shown in
Fig.\ref{fig1}. This lattice has a unit cell with three different lattice sites whose two sites ($A,C$) like in graphene form a honeycomb
lattice with hopping amplitude $t_{AC}=t_1$ and additional $B$ sites at the center of each hexagon are connected to the $C$ sites with
hopping amplitude $t_{BC}=t_2$. The two hopping parameters $t_1$ and $t_2$ are not equal in general. The corresponding model is known as the
$\alpha-\mathcal{T}_3$ model [\onlinecite{Raoux}]. The dice model corresponds to the limit $t_1=t_2$. The basis vectors of the
triangle Bravais lattice are
\begin{align}
\vec{a}_1=(\sqrt{3},\,\,0)a, \quad \vec{a}_2=\left(\frac{\sqrt{3}}{2},\,\,\frac{3}{2}\right)a,
\end{align}
where $a$ is the distance between two neighbors. The set of vectors with $\vec{a}_3=\vec{a}_2-\vec{a}_1$
\begin{align}
	\vec{\delta}_{1}=\frac{\vec{a}_1+\vec{a}_2}{3},\quad \vec{\delta}_{2}=\frac{\vec{a}_3-\vec{a}_1}{3}, \quad
	\vec{\delta}_{3}=-\frac{\vec{a}_2+\vec{a}_3}{3}
\end{align}
connect atoms from A sublattice with nearest C atoms (also these vectors with minus sign connect atoms B with C).
The tight-binding equations are [\onlinecite{Bercioux}]
\begin{align}
&\epsilon\Psi_C(\vec{r})=-t_1\sum\limits_{j}\Psi_{A}(\vec{r}+\vec{\delta}_{j})-t_2\sum\limits_{j}\Psi_{B}(\vec{r}-\vec{\delta}_{j}),
\nonumber\\
&\epsilon\Psi_{A}(\vec{r})=-t_1\sum\limits_{j}\Psi_{C}(\vec{r}-\vec{\delta}_{j}),\nonumber\\
&\epsilon\Psi_{B}(\vec{r})=-t_2\sum\limits_{j}\Psi_{C}(\vec{r}+\vec{\delta}_{j}).
\label{eqs:tight-binding}
\end{align}
 The corresponding Hamiltonian in momentum space reads [\onlinecite{Raoux}]
\begin{align}
\label{TB-Hamiltonian}
H=\left(\begin{array}{ccc}
0 & f_{\vec{k}}\cos\phi & 0\\
f^{*}_{\vec{k}}\cos\phi & 0 & f_{\vec{k}}\sin\phi\\
0 & f^{*}_{\vec{k}}\sin\phi & 0
\end{array}\right),
\quad \alpha \equiv \tan\phi=\frac{t_2}{t_1},\quad f_{\vec{k}}=-\sqrt{t_1^2+t_2^2}\,(1+e^{-i\vec{k}\vec{a}_2}+e^{-i\vec{k}\vec{a}_{3}}).
\end{align}
It is easy to find the energy spectrum of the above Hamiltonian, which is qualitatively the same for any $\alpha$ and consists of
three bands: a zero-energy flat band, $\epsilon_0(\mathbf{k})=0$, and two dispersive bands
\begin{equation}
\epsilon_\lambda(\mathbf{k})=\lambda|f_{k}|=\lambda \sqrt{t_1^2+t_2^2}\bigg[3+2(\cos(\vec{a}_1\vec{k})+\cos(\vec{a}_2\vec{k})+
\cos(\vec{a}_3\vec{k}))\bigg]^{1/2},\quad \lambda=\pm.
\end{equation}
The presence of a completely flat band with zero energy is perhaps one of the remarkable properties of the $\alpha-\mathcal{T}_{3}$ lattice
model. Since we will consider in this paper the boundary conditions for fermions in the dice model, we will set $\alpha=1$ in what
follows and denote $t_1=t_2=t/\sqrt{2}$.

There are six values of momentum for which $f_{\vec{k}}=0$ and all three bands meet. They are situated at the corners of the hexagonal
Brillouin zone. Two inequivalent point, for example, are
\begin{align}
	\vec{K}=\frac{2\pi}{a}\left(\frac{\sqrt{3}}{9},\,\frac{1}{3}\right),\quad \vec{K}'=\frac{2\pi}{a}\left(-\frac{\sqrt{3}}{9},\,
	\frac{1}{3}\right).
\end{align}
For momenta near the $K$-points, $\vec{k}=\vec{K}(\vec{K}')+\tilde{\vec{k}}$  we find that $f_{\vec{k}}$ is linear in $\tilde{\vec{k}}$,
i.e., $f_{\vec{k}}=\hbar v_F(\xi \tilde{k}_x-i\tilde{k}_y)$ with $\xi=\pm$, where $v_F=3ta/2\hbar$ is the Fermi velocity, and in what follows
we omit for the simplicity of notation the tilde over momentum. Thus, we obtain the low-energy Hamiltonian near the $K$-point in the
form [\onlinecite{Nicol}]
\begin{equation}\label{Hd-hamiltonian}
	\mathcal{H}_{d}=\hbar v_F\mathbf{S}\mathbf{k}=\frac{\hbar v_F}{\sqrt{2}}\left(\begin{array}{ccc}
	0 & k_{-} & 0\\
	k_{+} & 0 & k_{-}\\
	0 & k_{+} & 0
	\end{array}\right), \quad S_x=\frac{1}{\sqrt{2}}\left(\begin{array}{ccc}
	0 & 1 & 0\\
	1 & 0 & 1\\
	0 & 1 & 0
	\end{array}\right), \quad S_y=\frac{1}{\sqrt{2}}\left(\begin{array}{ccc}
	0 & -i & 0\\
	i & 0 & -i\\
	0 & i & 0
	\end{array}\right),
\end{equation}
where $\mathbf{S}$ are the spin matrices of the spin 1 representation and $k_{\pm}=k_x\pm ik_{y}$. The Hamiltonian acts on
three-component wave functions $\Psi^T=(\Psi_{A},\Psi_{C},\Psi_{B})$. In our analysis of
boundary conditions for different lattice terminations, we will need, in general, the full low-energy Hamiltonian in both $K$ and
$K^{\prime}$ valleys. Therefore, it is worth writing down this Hamiltonian explicitly

\begin{align}
	H=\left(\begin{array}{cc}
	\mathcal{H}_{d} & 0\\
	0 & -\mathcal{H}_{d}
	\end{array}\right),
\label{Diraclike-Hamiltonian}
\end{align}
which acts on 6-component spinor $\Psi=\left(\psi_{A},\psi_{C},\psi_{B}, \psi_{B}',\psi_{C}',\psi_{A}'\right)$, where like in graphene
the $A$ and $B$ spinor components are interchanged in the $K^{\prime}$ valley. We note that the tight-binding equations
(\ref{eqs:tight-binding}) have the electron-hole symmetry $\psi_C\rightarrow -\psi_C, \epsilon\rightarrow -\epsilon$ or, equivalently,
$\psi_A\rightarrow -\psi_A, \psi_B\rightarrow -\psi_B,\epsilon\rightarrow -\epsilon$. For the tight-binding Hamiltonian
(\ref{TB-Hamiltonian}), as well as the continuum Dirac-like Hamiltonian (\ref{Diraclike-Hamiltonian}),
this symmetry is translated into the anticommutation relation $\{H,\hat{C}\}=0$ with the charge conjugation
operator (\ref{C-operator}) in Appendix \ref{sec:M_parametric}. This particle-hole symmetry implies that if $E(\mathbf{k})$ is an eigenvalue
at given $\mathbf{k}$, then so is $-E(\mathbf{k})$. Since, in the present case, the total number of bands is odd, this makes it necessary the existence
of a zero eigenvalue at all $\mathbf{k}$, hence a zero energy flat band. The Hamiltonian \eqref{Diraclike-Hamiltonian} is invariant also with respect
to the time reversal $\hat{T}$ transformation where the operator $\hat{T}$ has the form $\hat{T}=\tau_1\otimes F\hat{K}$, here $\hat{K}$
is the operator of complex conjugation and the matrix $F$ is given by Eq.\eqref{F-matrix}.

In the analysis of boundary conditions, it is convenient to represent $6 \times 6$ matrices in the form of a tensor product
$\tau_{\mu} \otimes \lambda_{\nu}$, where $\tau_{\mu}=(\tau_0,\tau_i)$ and $\lambda_{\nu}=(\lambda_0,\lambda_j)$ act in the valley and
sublattice spaces, respectively. Here $\tau_i$ are the Pauli matrices, $\lambda_j$ are the Gell-Mann matrices, and $\tau_0$ and $\lambda_0$
are the unit $2\times2$ and $3\times3$ matrices, respectively. Clearly, if no leads are attached to the material, the electric current through
boundary should vanish. The current operator in the direction $\mathbf{n}$ normal to a boundary in our theory reads
\begin{align}\label{eq:current}
	\mathbf{n}\mathbf{J}=v_{F}\tau_3\otimes (\mathbf{S}\mathbf{n})=\frac{v_{F}}{\sqrt{2}}\tau_3\otimes\left(\begin{array}{ccc}
	0 & n_{-} & 0\\
	n_{+} & 0 & n_{-}\\
	0 & n_{+} & 0
	\end{array}\right), \quad\quad n_{\pm}=n_x \pm in_y.
\end{align}
Since the current operator is not a differential operator, vanishing of electric current at boundary cannot be formulated as the Neumann
condition as is usual in nonrelativistic physics. The same situation occurs in graphene whose low-energy Hamiltonian is also linear in momentum.
Therefore, like in graphene [\onlinecite{Falko,Akhmerov}], the general boundary condition for the current to vanish at boundary can be
formulated as a requirement that the wave function satisfies the following condition at boundary:
\begin{align}
	\Psi_{boundary}=M\Psi_{boundary},
\label{general-BC}
\end{align}
where matrix $M$ is Hermitian and anticommutes with the current operator, i.e.,
\begin{align}
\label{eq:general_condition_M}
	M=M^{\dagger}, \quad
	\{M, \mathbf{n}\mathbf{J}\}=0.
\end{align}
In view of $\langle\Psi|\mathbf{n}\mathbf{J}|\Psi\rangle=\langle\Psi|(\mathbf{n}\mathbf{J})M|\Psi\rangle=-\langle\Psi|M(\mathbf{n}\mathbf{J})|
\Psi\rangle=-\langle\Psi|\mathbf{n}\mathbf{J}|\Psi\rangle$, the anticommutation of $M$ with the current operator guarantees that the current
normal to the boundary vanishes. However, unlike graphene [\onlinecite{Akhmerov}] we cannot prove the inverse statement that the anticommutation
relation of $M$ with the current operator follows from the current conservation requirement because $\text{det}[\mathbf{n}\mathbf{J}]=0$ in the
case under consideration. The most general form of $6\times6$ matrix $M$ is considered in Appendix\ref{sec:M_parametric}.

\section{Ribbons with zigzag boundary conditions}
\label{sec:zigzag-simple}

\begin{figure}[h!]
    \centering
    \includegraphics[scale=0.35]{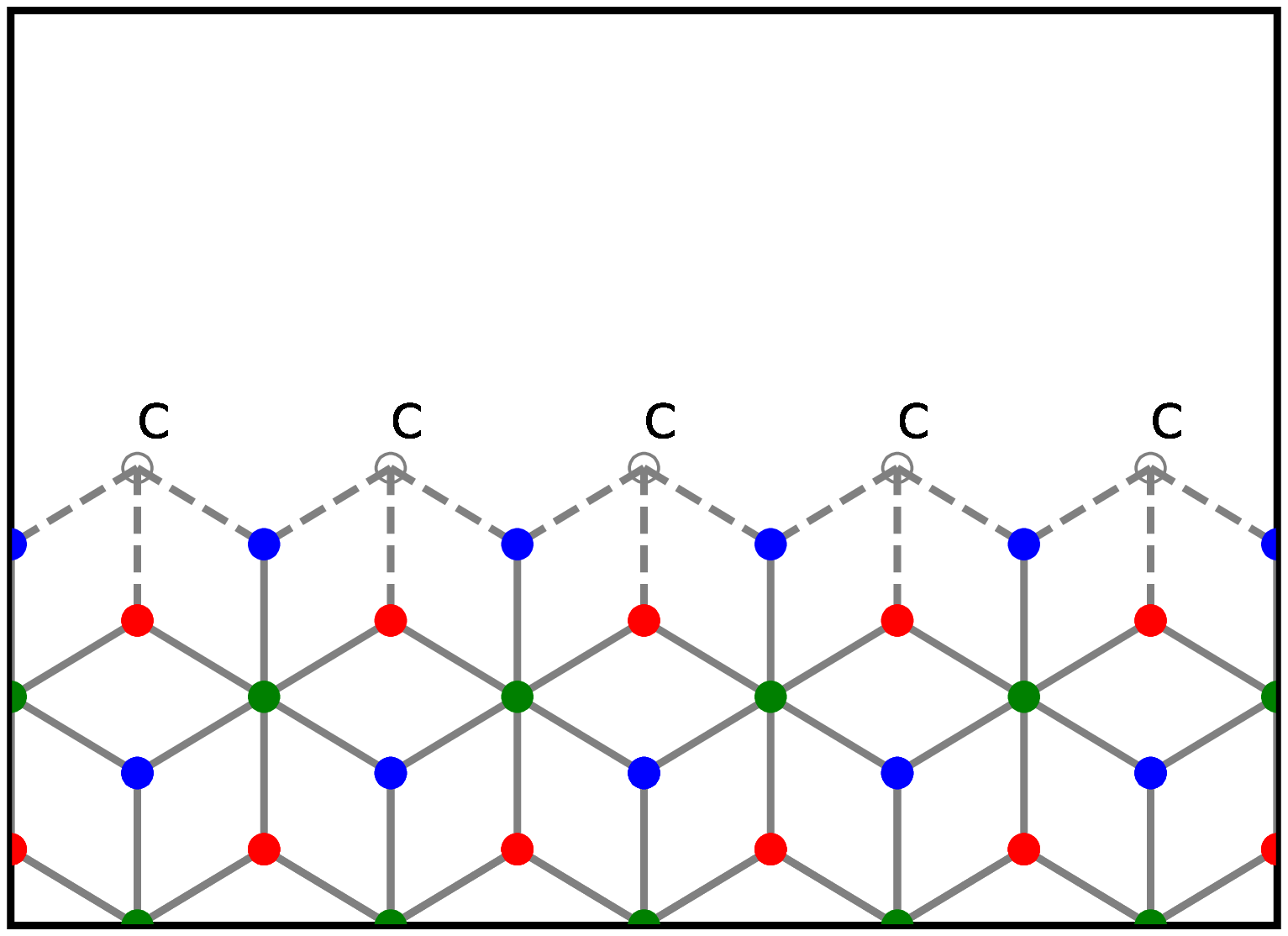}
    \includegraphics[scale=0.35]{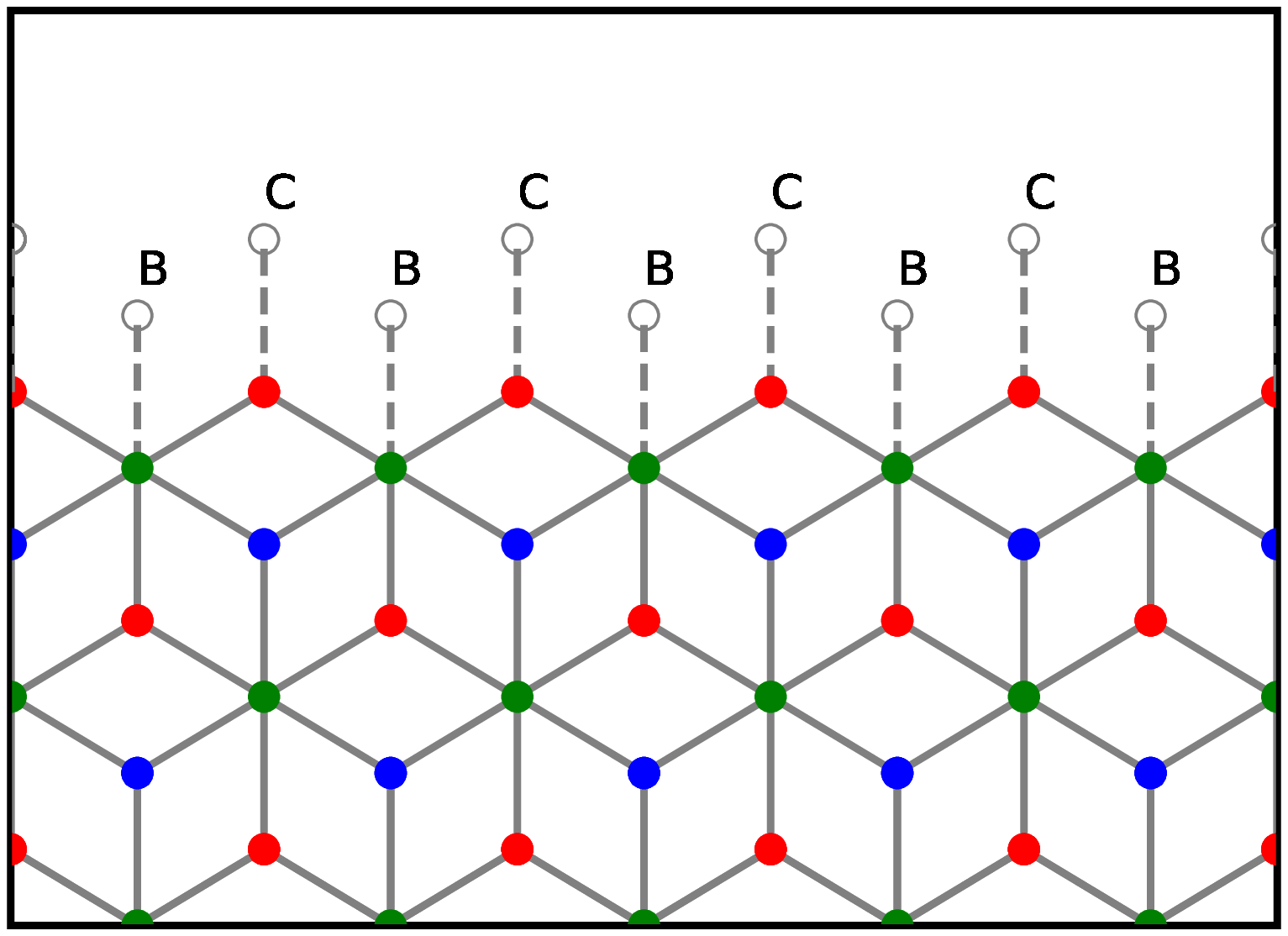}
    \includegraphics[scale=0.35]{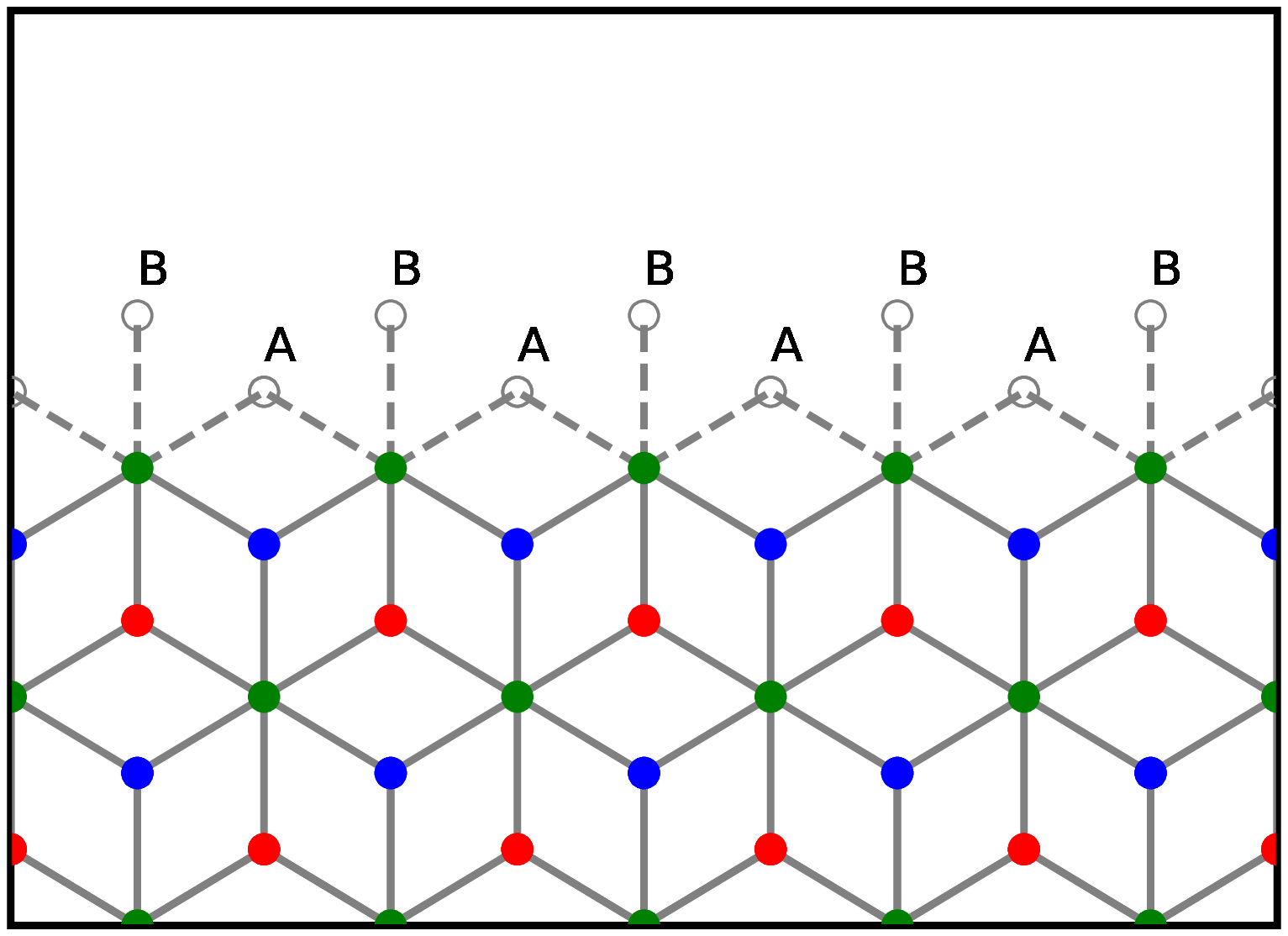}
    \includegraphics[scale=0.35]{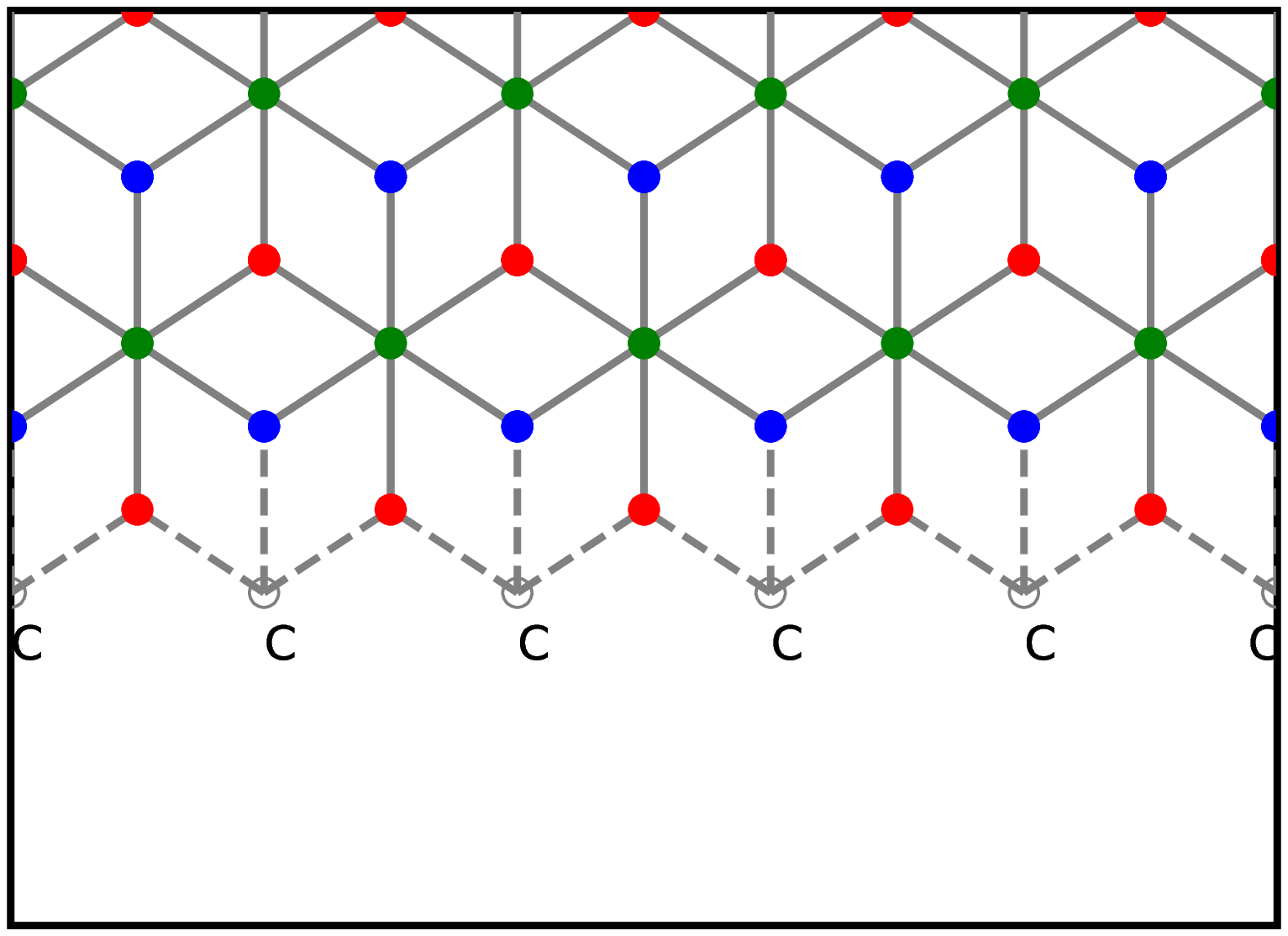}
    \includegraphics[scale=0.35]{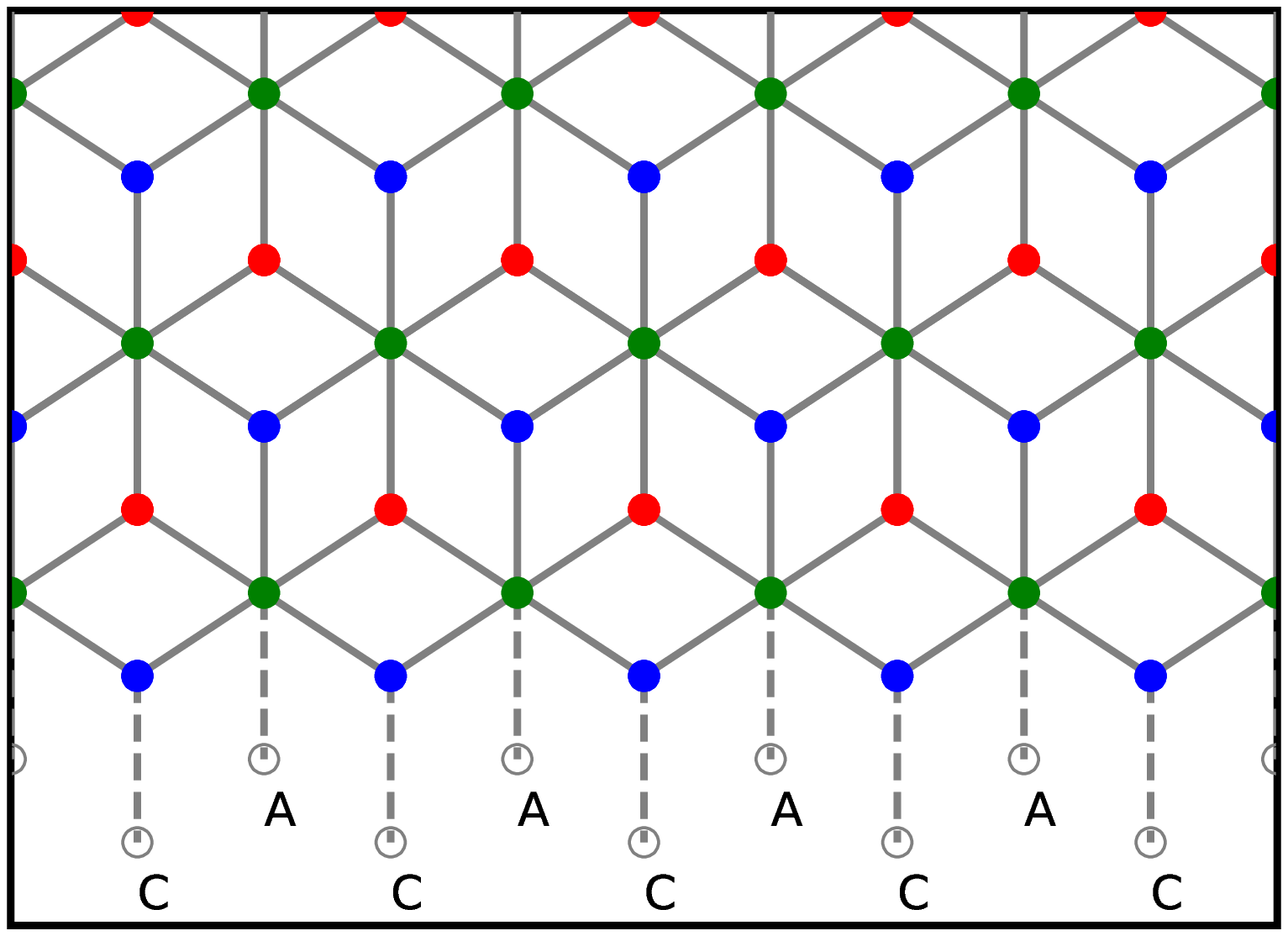}
    \includegraphics[scale=0.35]{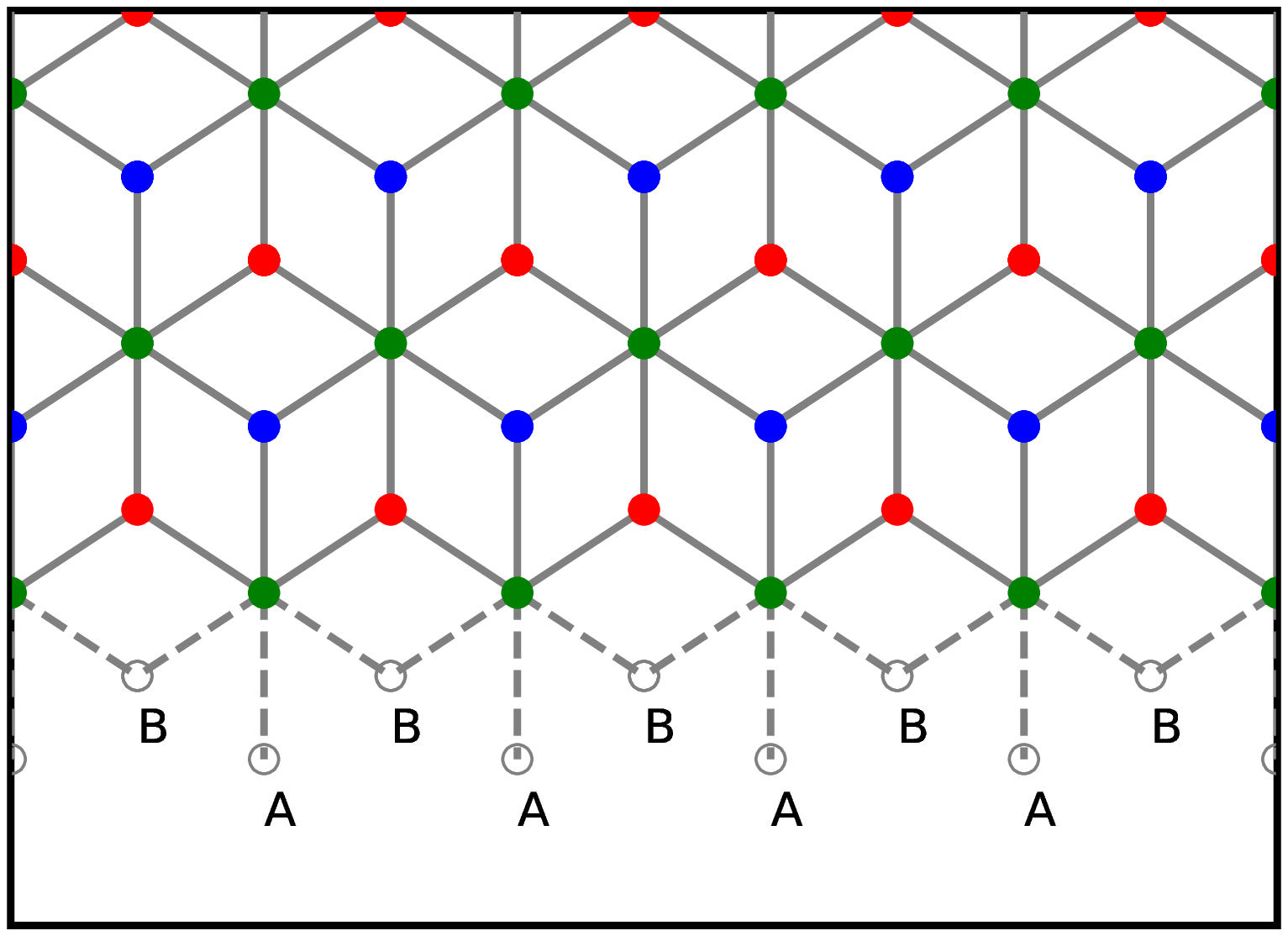}
    \caption{Upper panels: 3 possible types of the zigzag termination at $y=L$ with \textbf{C}, \textbf{CB}, and \textbf{BA} missing atoms.
    Lower panels: 3 possible types of zigzag termination at $y=0$ with \textbf{C}, \textbf{CA}, and \textbf{AB} missing atoms.
    }
    \label{fig2}
\end{figure}

In this section, we will study the boundary conditions and electron states in ribbons with zigzag edges along the $y=0$ and
$y=L$ sides. Since the dice lattice does not have mirror symmetry (or, equivalently, the $\pi$ rotational symmetry), possible
zigzag terminations on both lower and upper sides of a ribbon should be analyzed. The corresponding terminations are displayed in
Fig.\ref{fig2}. Its upper and lower panels imply that there are four possible zigzag terminations. It is worth recalling here that
graphene ribbons admit only one type of the zigzag edge.

Since the zigzag boundary conditions do not mix wave functions from different valleys, it suffices to perform our analysis in the $K$ valley
using the low-energy Hamiltonian \eqref{Hd-hamiltonian}. In view of the translation symmetry in the $x$-direction, we seek the wave
function in the form $\Psi_{\mu}=e^{ik_x x}\phi_{\mu}(y)$ and replace $k_y\to-i\p_y$. Then we obtain the system of equations for
$\phi_{\mu}(x)$ with $\mu=(A,C,B)$,
\begin{align}\label{kx+dy}
\left(\begin{array}{ccc}
0 & k_x-\p_y & 0\\
k_x+\p_y & 0 & k_x-\p_y\\
0 & k_x+\p_y & 0
\end{array}\right)\left(\begin{array}{c}
\phi_A\\
\phi_C\\
\phi_B
\end{array}\right)=\tilde{\epsilon}\left(\begin{array}{c}
\phi_A\\
\phi_C\\
\phi_B
\end{array}\right),
\end{align}
where $\tilde{\epsilon}=\frac{\epsilon \sqrt{2}}{\hbar v_F}$ and  $y$ belongs to the $[0, L]$ interval. For $\tilde{\epsilon}\neq 0$,
expressing $\phi_A$ and $\phi_B$ through $\phi_C$ and then substituting them into the second line of Eq.\eqref{kx+dy}, we obtain the
following second-order equation for $\phi_C$:
\begin{align}
\label{zigzag_eq:C-eigenproblem}
(\p_{y}^{2}-k_{x}^2)\phi_C=-\frac{\tilde{\epsilon}^2}{2}\phi_C,
\end{align}
whose general solution is given by (we use the short-hand notation $z=\sqrt{\frac{\tilde{\epsilon}^2}{2}-k_{x}^2}$\,)
\begin{align}
\label{zigzag:phi_C-sol}
\phi_C(y)=
A e^{izy}+B e^{-izy}.
\end{align}
Then we easily find the following expressions for the $\phi_A$ and $\phi_B$ components
\begin{eqnarray}
\label{zigzag:gensol_AB}
\left(\begin{array}{c}\phi_A\\ \phi_B\end{array}\right)=\frac{1}{\tilde{\epsilon}}\left(\begin{array}{c}(k_x-iz)Ae^{izy}+(k_x+iz)Be^{-izy}\\
(k_x+iz)Ae^{izy}+(k_x-iz)Be^{-izy}
\end{array}\right).
\end{eqnarray}
For $\tilde{\epsilon}=0$ the component $\phi_C=0$ and we get one equation for two functions,
\begin{equation}
(k_x+\partial_y)\phi_A+(k_x-\partial_y)\phi_B=0,
\label{eq:twofunctions}
\end{equation}
that leads to infinite degeneracy of this band. The above equations are analyzed below for different boundary conditions.

\subsection{Analytic results at non-zero energy}
\label{sec:zigzag-analytic}
In this subsection, we enumerate all possible zigzag terminations, analyze the corresponding boundary conditions, and determine
analytically the energy spectrum for ribbons with zigzag edges at non-zero energy.

\begin{enumerate}
\item The \textbf{C}-\textbf{C} boundary conditions.

It is very easy to check that vanishing of the $C$-component of the spinor wave function at the boundaries
$\phi_C(y=0)=\phi_C(y=L)=0$, where $L$ is the width of the ribbon, follows from the general boundary condition (\ref{general-BC})
with the matrix
     \begin{align}
    M_{C}=\tau_0\otimes\left(\begin{array}{ccc}
     1 & 0 & 0\\
     0 & -1 & 0\\
     0 & 0 & 1
     \end{array}\right).
      \end{align}
Note that the other necessary conditions (\ref{eq:general_condition_M}) on matrix $M$ are satisfied too. This matrix also preserves
both time reversal and electron-hole symmetries. Since $M_C$ does not mix states from different $K$ points, we omit below
the matrix $\tau_0$. Applying the obtained boundary conditions to the general solution \eqref{zigzag:phi_C-sol}, we easily find the spectrum
\begin{align}
\label{zigzag_eq:dispersion_of_modes}
\tilde{\epsilon}_n(k_x)=\pm\sqrt{2k_{x}^{2}+\frac{2\pi^2 n^2}{L^2}},\quad n=1,2,\dots
\end{align}
and the corresponding wave functions normalized in one valley as $\int\limits_0^Ldy\Psi^\dagger(k_x,y)\Psi(k_x,y)=1$,
\begin{align}
	\Psi_{n}(k_x,y)=\frac{1}{\tilde{\epsilon}_n\sqrt{L}}\left(\begin{array}{c}
	k_x\sin\left(\frac{\pi n y}{L}\right)-\frac{\pi n}{L} \cos\left(\frac{\pi n y}{L}\right)\\
	\tilde{\epsilon}_n\sin\left(\frac{\pi n y}{L}\right)\\
		k_x\sin\left(\frac{\pi n y}{L}\right)+\frac{\pi n}{L} \cos\left(\frac{\pi n y}{L}\right)
	\end{array}\right)e^{ik_x x},
\label{zigzageq:wavefunctions}
\end{align}
which describe the particle and hole bands for positive and negative energies, respectively. These are extended bulk states which are gapped
due to a spatial confinement in a finite width ribbon.

\item The \textbf{BA}-\textbf{AB} boundary conditions.

It is straightforward to satisfy the boundary conditions $\phi_{A}(y=0)=\phi_{B}(y=0)=0$ and $\phi_{A}(y=L)=\phi_{B}(y=L)=0$
by choosing the matrix $M$ in the form
	\begin{align}
		M_{AB}=\left(\begin{array}{ccc}
		-1 & 0 & 0\\
		0 & 1 & 0\\
		0 & 0 & -1
		\end{array}\right).
		\label{AB-naive}
	\end{align}
Obviously, conditions (\ref{eq:general_condition_M}) are satisfied also because $M_{AB}=-M_{C}$ and $M$ defined in
Eq.(\ref{AB-naive}) give four linearly independent boundary conditions on the components of the wave function. Then combining them
with Eq.\eqref{zigzag:gensol_AB}, we obtain only trivial solutions.  However, the direct numerical tight-binding calculations in the
lattice model give nontrivial solutions shown in the left panel of Fig.\ref{fig4} (see also the corresponding discussion
in Sec.\ref{subsec:zigzag-numeric}). This means that we should try to find other \textbf{BA}-\textbf{AB} boundary conditions in the
continuum model which reproduce at low energies the numerical solutions found in the lattice model.

According to Eq.(\ref{normal-current-zigzag-1}) in Appendix \ref{appendix:zero-current}, the normal component of the current
vanishes if either $\phi_C=0$ or $\phi_A-\phi_B=0$ as is clear from Eq.(\ref{normal-current-zigzag-2}). Definitely, we should choose
the second variant because the first describes the case of \text{C} missing atoms considered above. Note that the boundary condition
$\phi_A-\phi_B=0$ is not just a lattice termination, but allows for local electric fields and strained bonds. For the equation
$\phi_A-\phi_B=0$, the corresponding matrix $M$ has the form
\begin{align}
\overline{M}_{AB}=\left(\begin{array}{ccc}
0 & 0 & 1\\
0 & 1 & 0\\
1 & 0 & 0
\end{array}\right).
\end{align}
Obviously, this matrix anticommutes with the $J_y$ current operator and preserves both T- and C-symmetries. Although
$\overline{M}_{AB}$ is quite different from $M_C$, the results in the cases of the \textbf{C}-\textbf{C} and \textbf{BA}-\textbf{AB} boundary
conditions are similar. Using solutions \eqref{zigzag:gensol_AB} and imposing the boundary conditions with the matrix $\overline{M}_{AB}$, we
obtain equations for constants $A$ and $B$. This gives the same spectrum as in the \textbf{C}-\textbf{C} zigzag ribbons with the
normalized wave functions
	\begin{align}
	\Psi_{n}(k_x,y)=\frac{1}{\tilde{\epsilon}_n\sqrt{L}}\left(\begin{array}{c}
	k_x\cos\left(\frac{\pi n y}{L}\right)+\frac{\pi n}{L} \sin\left(\frac{\pi n y}{L}\right)\\
	\tilde{\epsilon}_n\cos\left(\frac{\pi n y}{L}\right)\\
	k_x\cos\left(\frac{\pi n y}{L}\right)-\frac{\pi n}{L} \sin\left(\frac{\pi n y}{L}\right)
	\end{array}\right)e^{ik_x x},\quad n=0,1,2,\dots
	\end{align}
(compare these functions with those in Eq.(\ref{zigzageq:wavefunctions}).
Note that the solution with $n=0$ is special with the gapless linear energy dispersion $\tilde{\epsilon}=\pm \sqrt{2}k_x$ and constant wave
function $\phi_C(y)=const\neq 0$. This is the only case of ribbons with zigzag terminations which have bulk gapless (metallic) modes.
Such modes are absent for graphene zigzag ribbons [\onlinecite{Brey}].

\item The \textbf{C}-\textbf{AB} boundary conditions correspond to $\phi_C(y=0)=0$ and $\phi_A(y=L)-\phi_B(y=L)=0$. Combining
equations \eqref{zigzag:phi_C-sol} and \eqref{zigzag:gensol_AB}, we obtain the energy spectrum
\begin{align}\label{eq:AB_C-spectrum}
\tilde{\epsilon}^2_n(k_x)=2k_{x}^{2}+\frac{2\pi^2}{L^2}\left(n+\frac{1}{2}\right)^2, \quad n=0,1,2,\dots,
\end{align}
and wave functions
\begin{align}
	\Psi_{n}(k_x,y)=\frac{1}{\tilde{\epsilon}_n\sqrt{L}}\left(\begin{array}{c}
	k_x\sin\left(\frac{\pi  y}{L}\left(n+\frac{1}{2}\right)\right)-\frac{\pi }{L}\left(n+\frac{1}{2}\right) \cos\left(\frac{\pi y}{L}
\left(n+\frac{1}{2}\right)\right)\\
	\tilde{\epsilon}_n\sin\left(\frac{\pi  y}{L}\left(n+\frac{1}{2}\right)\right)\\
	k_x\sin\left(\frac{\pi  y}{L}\left(n+\frac{1}{2}\right)\right)+\frac{\pi }{L}\left(n+\frac{1}{2}\right) \cos\left(\frac{\pi y}{L}
\left(n+\frac{1}{2}\right)\right)
	\end{array}\right)e^{ik_x x}.
\end{align}

Obviously, spectrum (\ref{eq:AB_C-spectrum}) is shifted compared to that in Eq.(\ref{zigzag_eq:dispersion_of_modes}) due to
the presence of $1/2$ in the brackets and is plotted in the right panel of Fig.\ref{fig4}.

The analysis of the \textbf{BA}-\textbf{C} boundary conditions $\phi_A(y=0)-\phi_B(y=0)=0$, $\phi_C(y=L)=0$ is similar because it does not
matter to which side the \textbf{AB} and \textbf{C} boundary conditions are imposed.

\item The \textbf{CB}-\textbf{CA} boundary conditions.

Naively, one may try to use the boundary conditions $\phi_A(y=0)= \phi_C(y=0)=0$ and $\phi_B(y=L)= \phi_C(y=L)=0$. However,
the eigenvalue problem \eqref{kx+dy} becomes overdetermined for these boundary conditions and does not have nontrivial solutions.
Like in the case of the \text{BA}-\text{AB} boundary conditions considered above, the numerical tight-binding
calculations in the lattice model give nontrivial solutions shown in the right panel of Fig.\ref{fig3}. Once again, this
means that we should try to find other \textbf{CB}-\textbf{CA} boundary conditions in the continuum model which reproduce at low energies
the numerical solutions found in the lattice model.

Recall that Eq.(\ref{normal-current-zigzag-2}) in Appendix \ref{appendix:zero-current} implies the normal component of the current
vanishes if either $\phi_C=0$ or $\phi_A-\phi_B=0$. Since we have already used the second variant for the \textbf{BA}-\textbf{AB}
boundary conditions, the only remaining way to impose the boundary condition in the continuum theory is to use the condition
$\phi_{C}|_{y=L}=0$ as an approximation. Certainly, the corrections from the boundary conditions with the missing \textbf{A}
and \textbf{B} atoms in the lattice model may become notable at high energies. However, as we checked in Subsec.\ref{subsec:zigzag-numeric}
below, this is not important in the low-energy model. Therefore, the zigzag boundary conditions \textbf{CB}-\textbf{CA} in the low-energy
continuum model are similar to the \textbf{C}-\textbf{C} zigzag ones.

\item There are four other possible zigzag \textbf{C}-\textbf{CA}, \textbf{CB}-\textbf{C}, \textbf{CB}-\textbf{AB}, \textbf{BA}-\textbf{CA}
terminations of a ribbon, however, all of them are equivalent to the cases discussed above.

\end{enumerate}

Thus, we end up with the two main types of zigzag terminations \textbf{C} and \textbf{AB} on each side on a ribbon leading, obviously,
to four possible zigzag edges. Note that the spectrum in each case differs from that in graphene [\onlinecite{Brey}]
and there are no states localized near the edges of the ribbon. On the other hand, ribbons with the
\textbf{BA}-\textbf{AB} boundary conditions contain solutions of metallic type in bulk and this is a new feature of zigzag boundary
conditions in the dice model compared to graphene ribbons where metallic states in bulk are absent. [It is worth mentioning that
the dispersion relations for bulk states in graphene ribbons are essentially nonlinear unlike the bulk states in the dice lattice
model found here.] Ribbons with other combinations of terminations are insulators at zero chemical potential.

\subsection{Zero energy}
\label{sec:espilon=0_zigzag}

The case of zero energy is of a special interest. The crucial question is whether the zero energy flat band present in an infinite
size system survives in the presence of boundaries. It is appropriate to recall that the zero-energy solution in the dice model in the
absence of boundaries have $\phi_C\equiv 0$ [\onlinecite{Bercioux,Raoux}]. For the strip of finite width we also have $\phi_C\equiv 0$,
and  only one equation (\ref{eq:twofunctions}) for two components $\phi_A$, $\phi_B$ that reflects an infinite degeneracy of the
zero-energy band. An arbitrary function defined on a segment $[0,L]$ can be parameterized by the coefficients of its Fourier series.
Therefore, we seek the solutions of Eq.(\ref{eq:twofunctions}) in the form
\begin{align}
	\phi_A(y)=A_1 \sin(zy)+A_2\cos(zy), \quad \phi_B(y)=B_1\sin(zy)+B_2\cos(zy)
\end{align}
that gives the equation
\begin{align}\label{eq:zigzag-zero-sin-cos}
	(k_xA_1-zA_2+k_xB_1+zB_2)\sin(zy)+(zA_1+k_xA_2-zB_1+k_xB_2)\cos(zy)=0,
\end{align}
{which is identically satisfied for any $0<y<L$ when the coefficients near $\sin(zx)$ and $\cos(zx)$ are zero.

As was discussed in previous section, there are two main different types of conditions - \textbf{C} and \textbf{AB}. For brevity,
we analyze one of the possible terminations in Sec.\ref{sec:zigzag-analytic}, namely, the \textbf{BA}-\textbf{AB} termination with the
boundary conditions $\phi_A-\phi_B=0$. Equation \eqref{eq:zigzag-zero-sin-cos} and the boundary conditions at the $y=0$ and $y=L$ edges
give the system
\begin{eqnarray}
\label{eq:zigzag-zero-conditions}
k_x A_1-z A_2+k_x B_1+z B_2=0, \quad z A_1+k_x A_2-z B_1+k_x B_2=0,\nonumber\\
A_2-B_2=0,\quad (A_1-B_1)\sin(z L)+(A_2-B_2)\cos(z L)=0,
\end{eqnarray}
which has nontrivial solutions when $\sin(zL)=0$, i.e., $z=z_n=\frac{\pi n}{L}$ with $n=1,2\dots$. The normalized wave functions are
\begin{align}
	\Psi_0(z_n, k_x)=\frac{1}{\sqrt{L}}\left(k_x^2+\frac{\pi^2 n^2}{L^2}\right)^{-1/2}\left(\begin{array}{c}
	k_x \sin\left(\frac{\pi n y}{L}\right)-\frac{\pi n}{L} \cos\left(\frac{\pi n y}{L}\right)\\
	0\\
-k_x \sin\left(\frac{\pi n y}{L}\right)-\frac{\pi n}{L} \cos\left(\frac{\pi n y}{L}\right)
	\end{array}\right)e^{ik_x x}.
\end{align}
Solutions for other terminations can be found similarly. The found solutions are in accordance with the general solution of
the tight-binding Hamiltonian for the zero-energy band in the case of infinite system [\onlinecite{Bercioux,Raoux}].

\subsection{Numerical results}
\label{subsec:zigzag-numeric}

\begin{figure}
	\centering
	\includegraphics[scale=0.4]{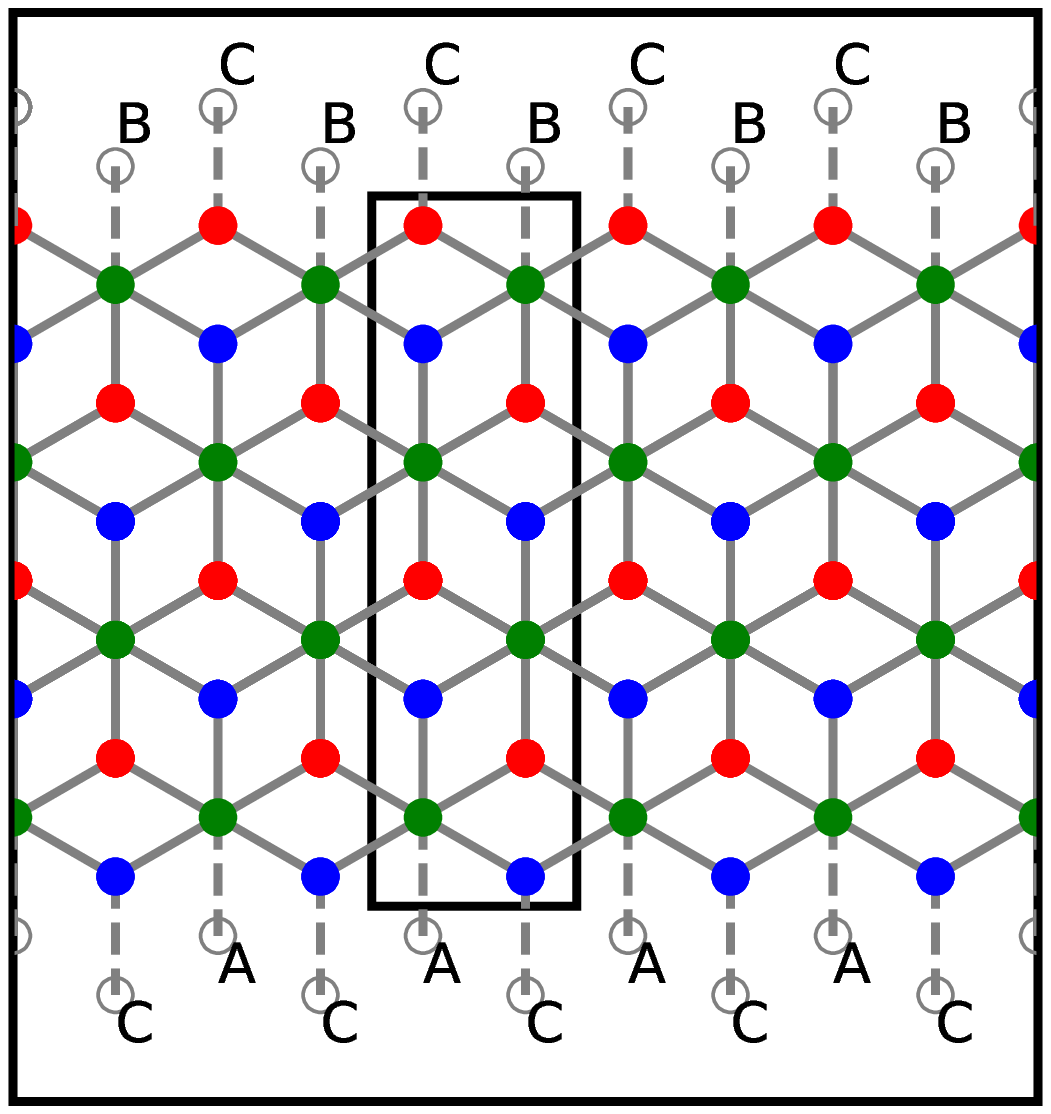}
	\includegraphics[scale=0.4]{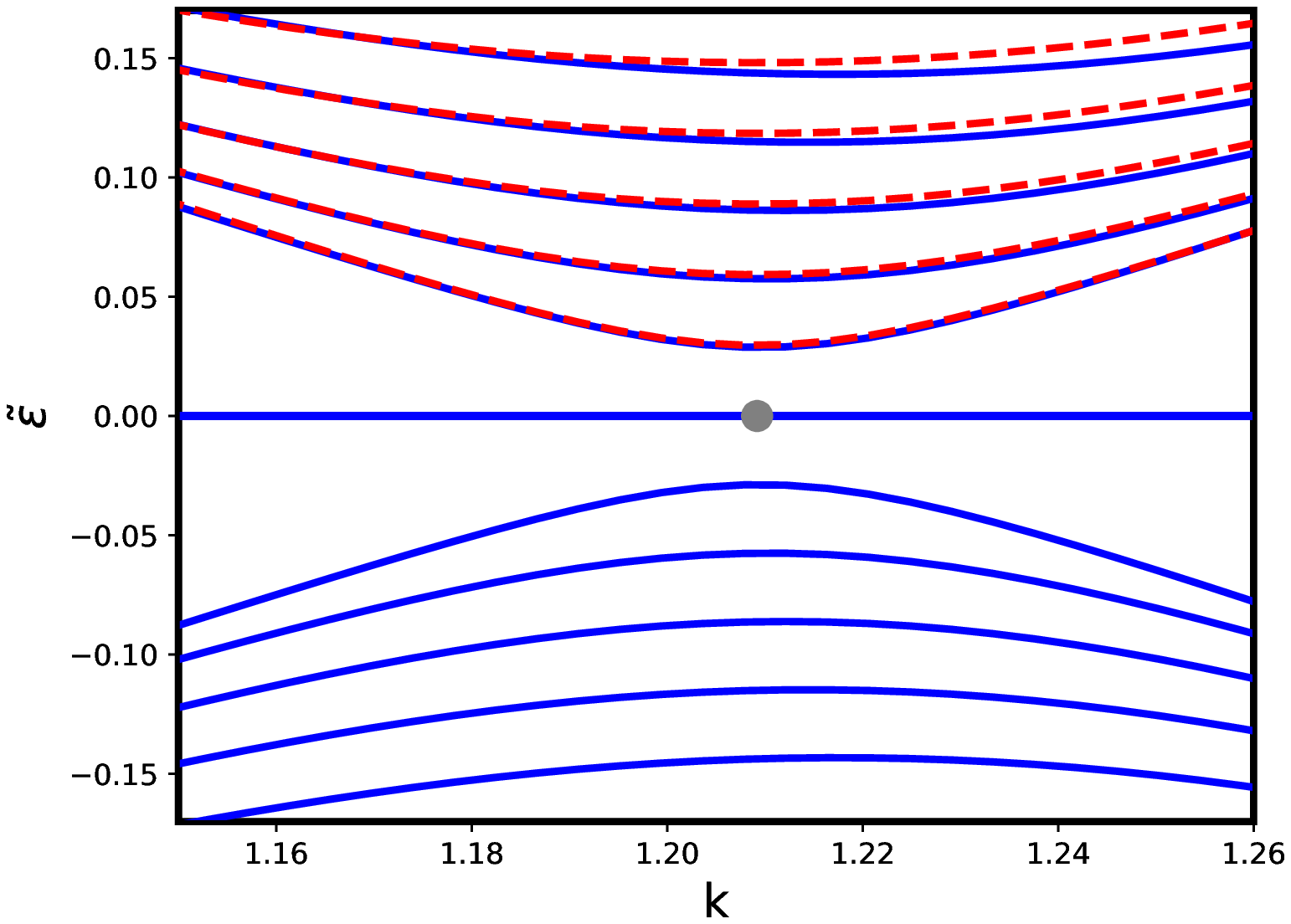}
	\caption{(Left panel) The cell denoted by black rectangle is used in the calculation of ribbons with the zigzag \textbf{CB}-\textbf{CA}
	boundary conditions. (Right panel) The energy bands for ribbons with \textbf{CB}-\textbf{CA} boundary conditions as functions of
    the wave vector $k$ parallel to the nanoribbon edge, measured with respect to the center of the Brillouin zone. The blue lines denote
    the energy levels determined from the tight-binding equations and the red dashed lines are plotted by using the theoretical formula
	\eqref{zigzag_eq:dispersion_of_modes} only at positive energy for the clarity of presentation. The gray point denotes the
	$K$-point. The number of elementary cells in the numerical calculations is 100.}
	\label{fig3}
\end{figure}

In the previous two subsections, we determined the energy spectrum and wave functions for ribbons with the zigzag boundary conditions
in the low energy continuum model. However, we met some problems in imposing the \textbf{BA}, \textbf{CB}, and \text{CA} boundary
conditions. Therefore, it is necessary to perform the calculations in the tight-binding model, compare the corresponding results,
and find out how the spectrum looks like at high energy where the low energy continuum model is, strictly speaking, not applicable.
For the unit computation cell with the zigzag \textbf{CB}-\textbf{CA} boundary conditions shown in left panel of Fig.\ref{fig3},
we plot in the right panel of the same figure the corresponding energy levels calculated at the $K$ point as well as the energy
levels in the low energy continuum model shown by red dashed lines. The latter are shown only in the upper energy half-plane for the
clarity of presentation because the energy levels in the lower half-plane trivially follow from the particle-hole symmetry of the spectrum.
Since the energy levels for the \textbf{C}-\textbf{C} and \textbf{CA}-\textbf{C} boundary conditions are practically indistinguishable
from the energy levels in the right panel of Fig.\ref{fig3}, we do not plot them separately. In addition, the left and right panels in
Fig.\ref{fig4} describe the results obtained for the \textbf{BA}-\textbf{AB} and \textbf{BA}-\textbf{C} boundary conditions, respectively.
Our main results are the following:
\begin{enumerate}
	\item The results found in the low energy continuum model are very accurate and the energy of the $n-$th level for $k=0$ equals
	$\tilde{\epsilon}_n=\frac{\sqrt{2}\pi n}{L}$, where $L=3Na/2$ is the width of the ribbon and $N$ is the number of elementary cells in
     the calculation cell.
	\item The \textbf{CB}-\textbf{CA} boundary conditions as well as the \textbf{C}-\textbf{CA} and \textbf{CB}-\textbf{C} ones lead to
	the same dispersion as the \textbf{C}-\textbf{C} boundary conditions. This supports our conclusion that at small momenta the \textbf{CB}
    and \textbf{CA} zigzag boundary conditions type are similar to the \textbf{C} boundary condition. However, the energy dispersion for the \textbf{BA}-\textbf{AB} boundary conditions shown on the left panel in Fig.\ref{fig4} is qualitatively different and contains two gapless
    modes.
	\item The \textbf{BA}-\textbf{C} boundary conditions lead to a shifted spectrum, as predicted by Eq.\eqref{eq:AB_C-spectrum}.
	\item The numerical results shown in Fig.\ref{fig5} demonstrate the energy levels found in the continuum and tight-binding
	models in the zigzag ribbons with \text{C}-\text{C}, \textbf{CB}-\textbf{CA}, and \textbf{BA}-\textbf{AB} boundary conditions throughout
    the Brillouin zone.
\end{enumerate}

\begin{figure}[h!]
	\centering
    \includegraphics[scale=0.4]{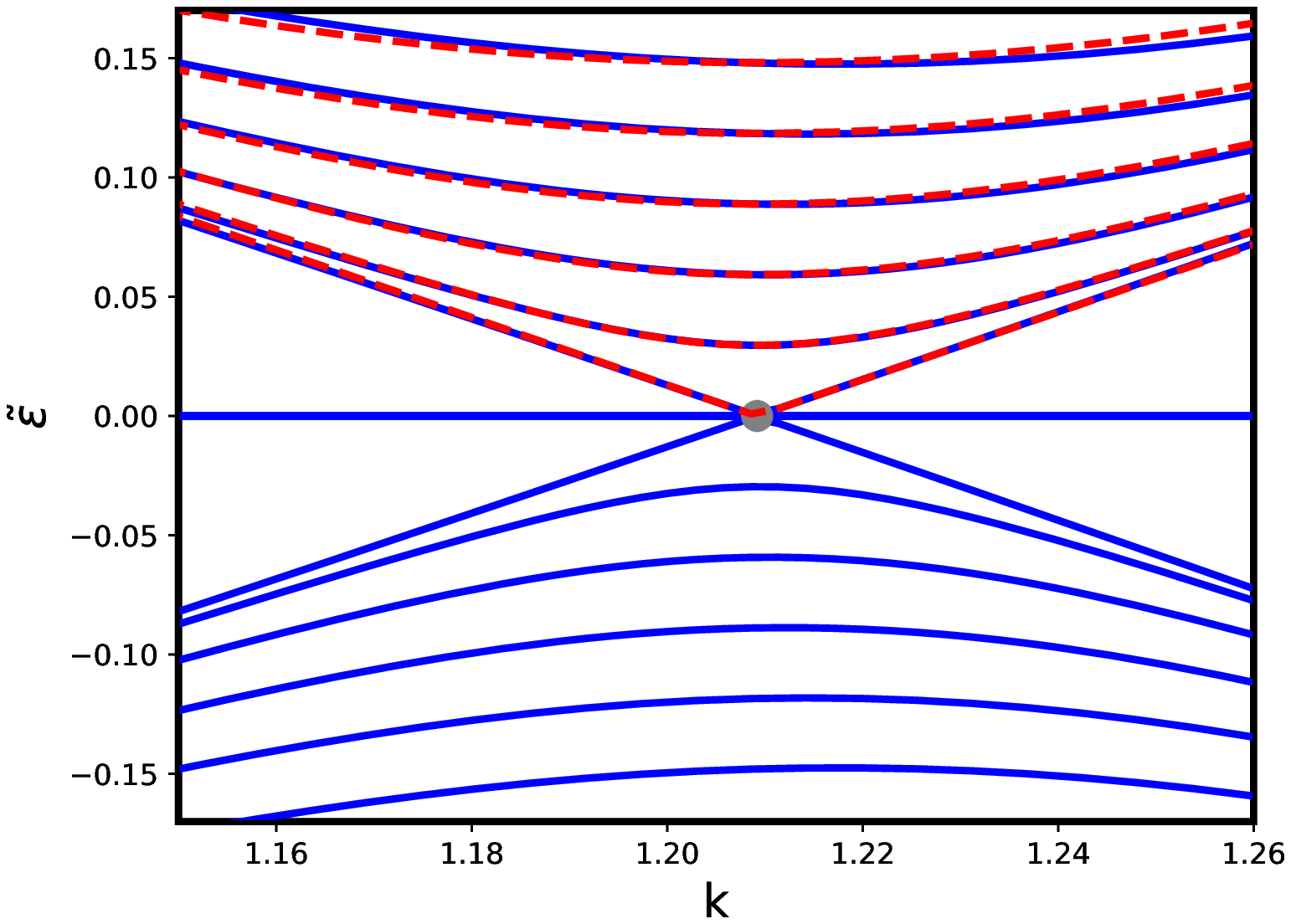}
	\includegraphics[scale=0.4]{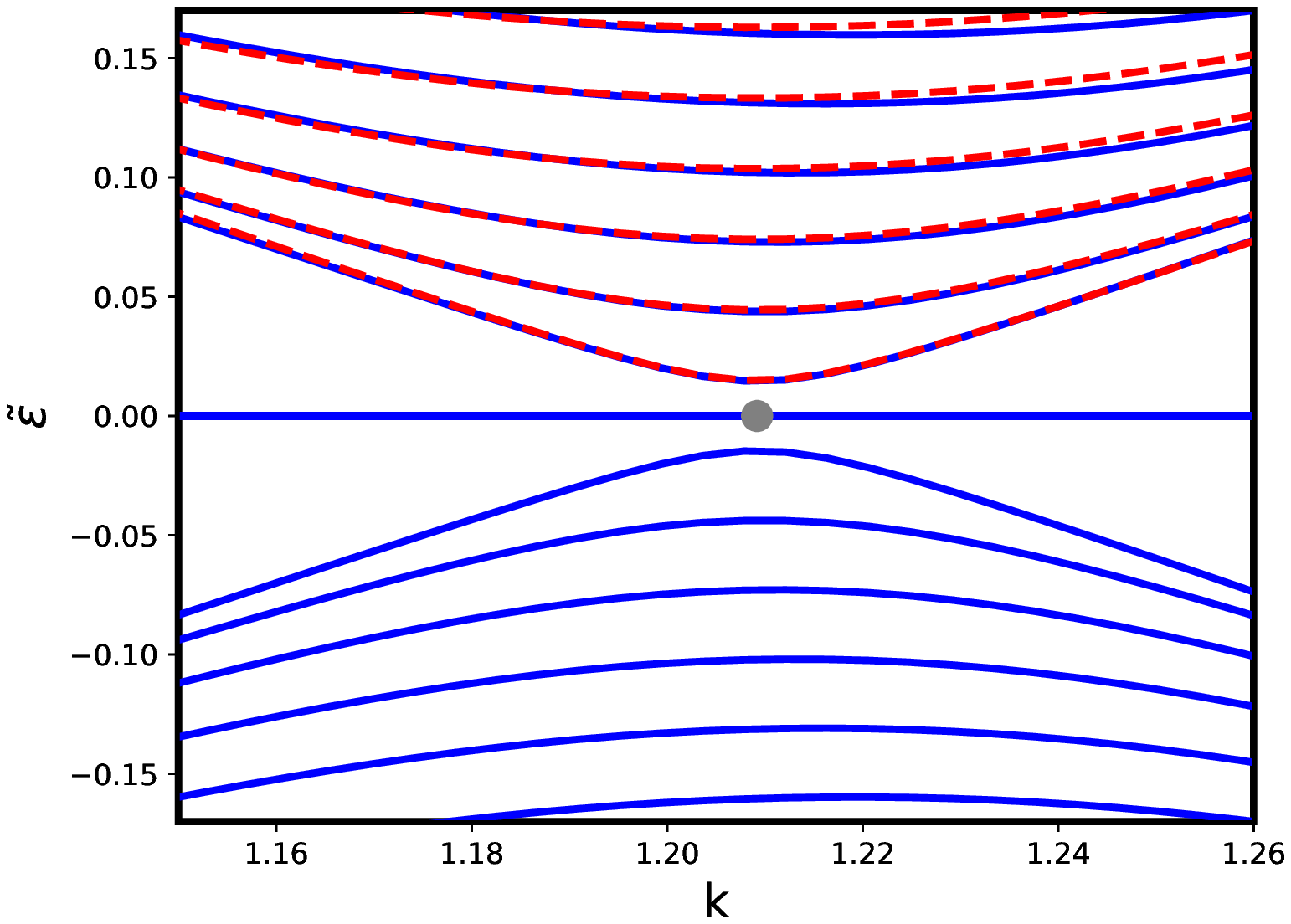}
	\caption{The left and right panels describe the results obtained for the \textbf{BA}-\textbf{AB} and \textbf{BA}-\textbf{C} boundary
	conditions, respectively. The number of elementary cells in the numerical calculations is 100.  Note that there are two
	gapless states for a ribbon with the \textbf{BA}-\textbf{AB} boundary conditions. The theoretical curves are represented as red dashed
    lines only in the upper energy half-plane for the clarity of presentation. The gray point denotes the $K$-point.}
	\label{fig4}
\end{figure}

\begin{figure}[h!]
	\centering
	\includegraphics[scale=0.35]{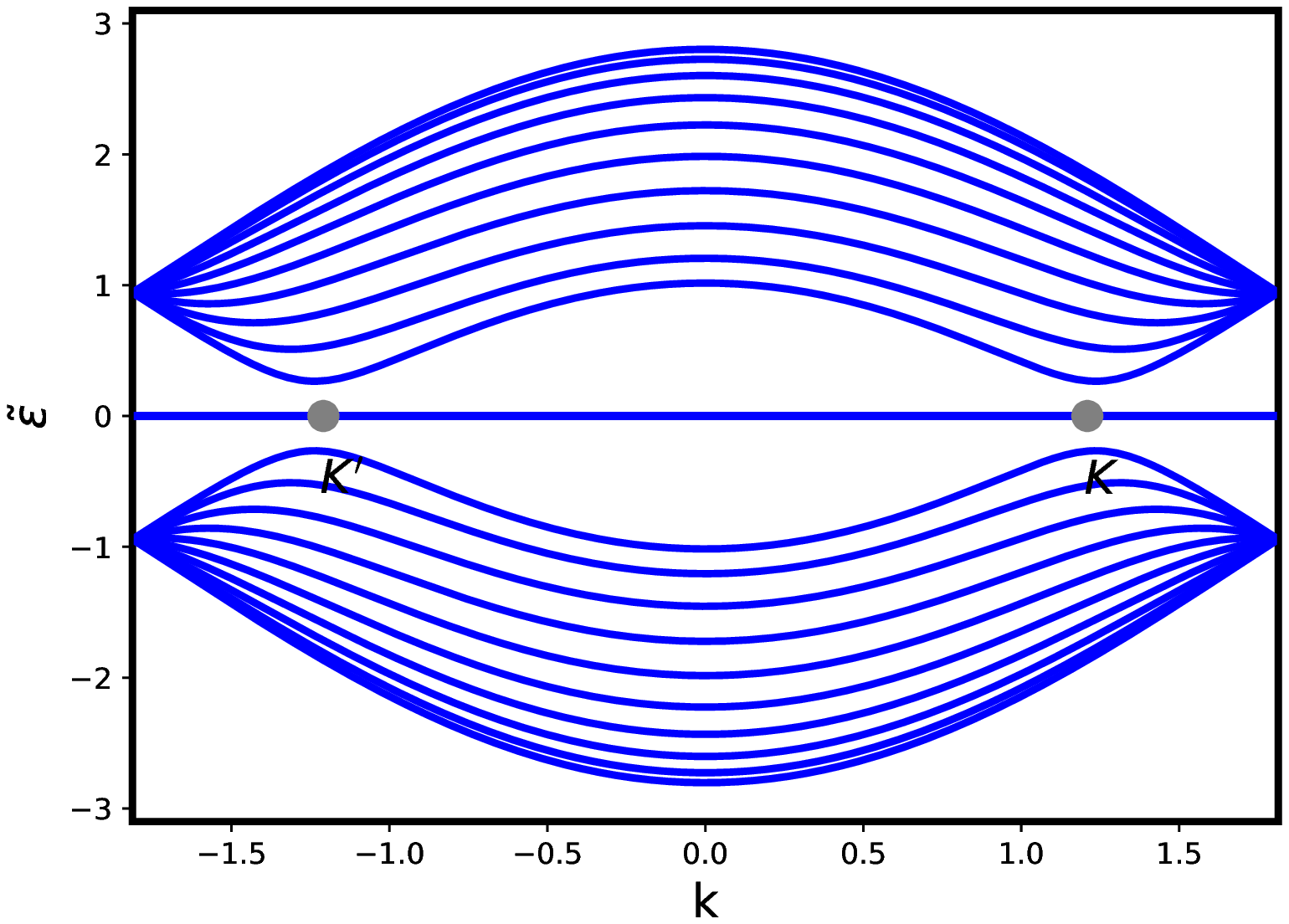}
	\includegraphics[scale=0.35]{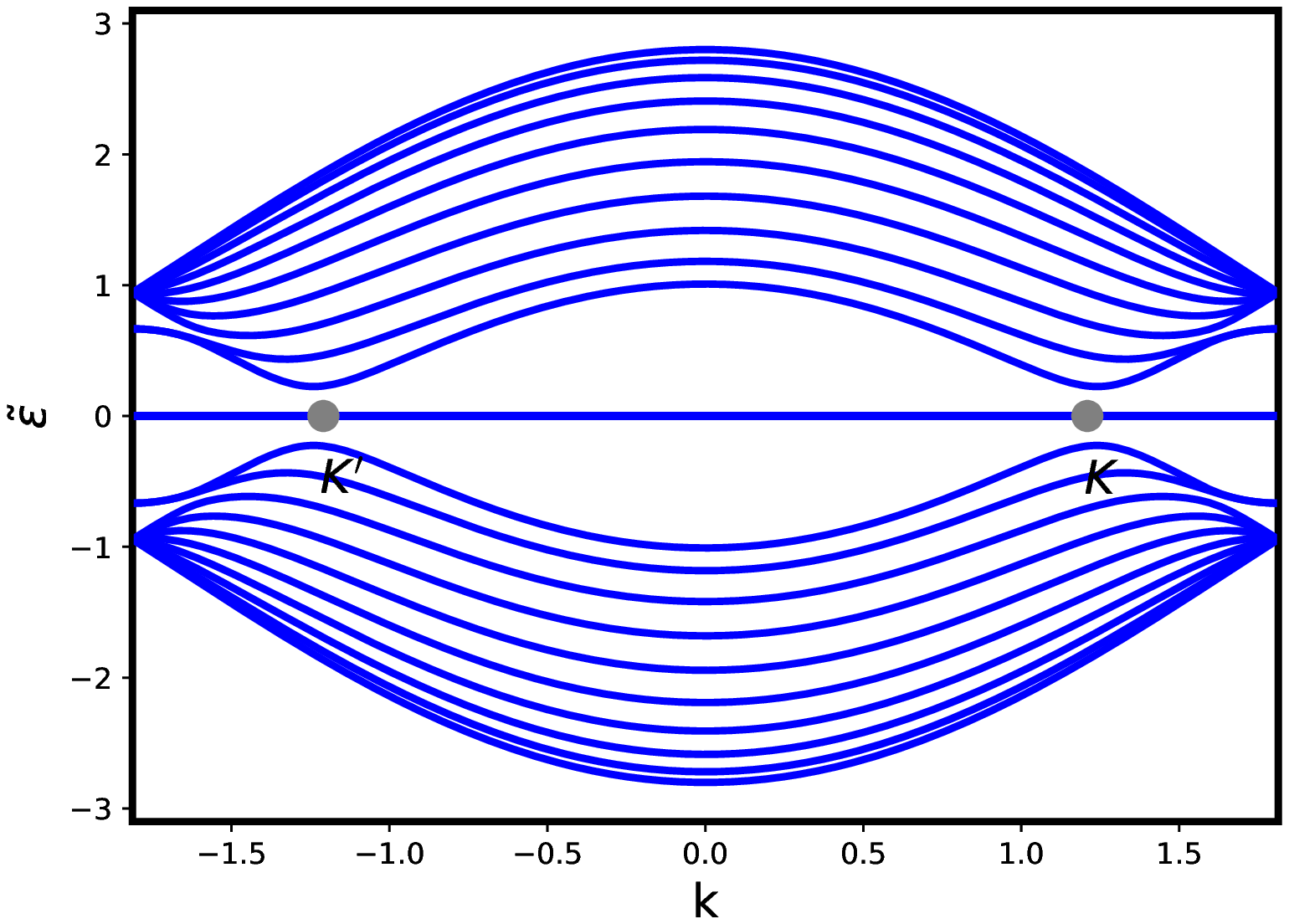}	
	\includegraphics[scale=0.35]{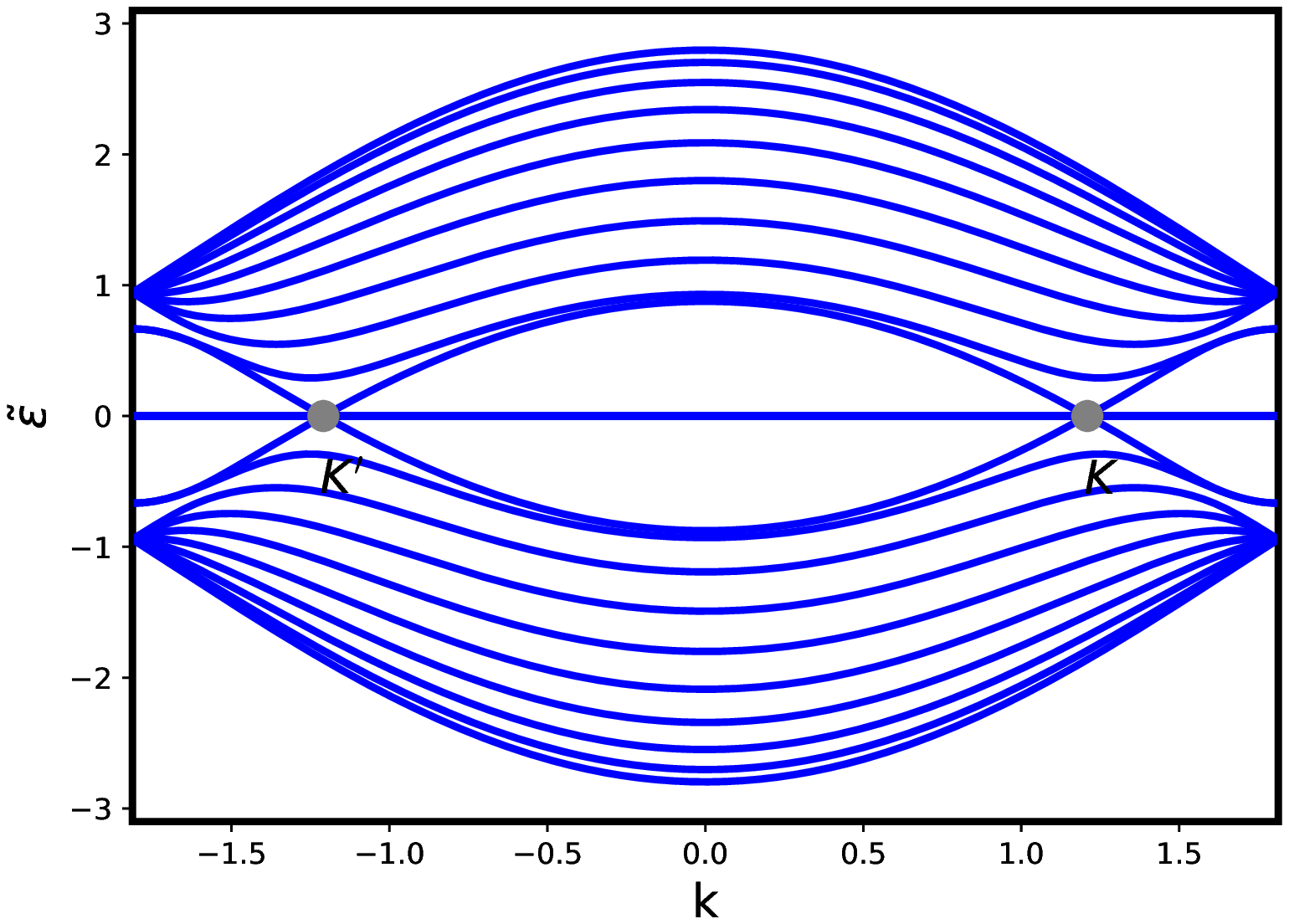}
	\caption{The numerically calculated energy spectrum throughout the Brillouin zone for ribbons with
			the zigzag \textbf{C}-\textbf{C} boundary conditions (left panel), the \textbf{CB}-\textbf{CA} boundary conditions (middle
			panel), and the \textbf{BA}-\textbf{AB} boundary conditions (right panel).
			The number of elementary cells in the calculations is 10.}
	\label{fig5}
\end{figure}

\section{Ribbons with armchair boundary conditions}
\label{sec:armchair}

In this section, we will study the electron states and energy spectrum in ribbons with the armchair boundary conditions imposed at
the $x=0$ and $x=L$ sides. Such a ribbon is schematically shown in Fig.\ref{fig6}. Following the derivation of corresponding boundary
conditions in graphene [\onlinecite{GeimRMP2009}], we find that the $\mu=A,\,B, C$ components of the wave function obey the equations
\begin{align}
	&\phi_{\mu}(x=0)=\phi_{\mu}'(x=0),\quad
	\phi_{\mu}(x=L)=e^{i\Delta KL}\phi_{\mu}'(x=L).
	\label{armchair-BC}
\end{align}
The armchair boundary conditions mix states from the different $K$ and $K^{\prime}$ valleys and the factor $\Delta K=\frac{4\pi}{3\sqrt{3}a}$
comes from the scalar product $(\vec{K}-\vec{K}^{\prime})(L\vec{e}_x)$, which describes the phase difference between states from different
valleys on the $x=L$ edge. Note that the phase in the second equation (\ref{armchair-BC}) is similar to graphene
[\onlinecite{Brey}]. Therefore, the matrix $M$ has nonzero off-diagonal blocks and equals
\begin{align}
\label{M-matrices-armchair}
	M_{1}\bigg|_{x=0}=\tau_1\otimes\left(\begin{array}{ccc}
    0 & 0 & 1\\
    0 & 1 & 0\\
    1 & 0 & 0
	\end{array}\right)=\tau_1\otimes F,\quad  	M_{2}\bigg|_{x=L}=\left(\begin{array}{cc}
	0 & e^{i\Delta K L}\\
	e^{-i\Delta K L} & 0
	\end{array}\right)\otimes\left(\begin{array}{ccc}
	0 & 0 & 1\\
	0 & 1 & 0\\
	1 & 0 & 0
	\end{array}\right).
\end{align}
Obviously, $M^{\dagger}_{1,2}=M_{1,2}$ and the matrices $M_{1,2}$ anticommute with the normal component of the current
$\{M_{1,2},\mathbf{n}\mathbf{J}\}=0$ for $n_x=\pm1$. Both matrices $M_{1,2}$ also preserve $T$- and $C$-symmetries. Our next step is to find nontrivial solutions for ribbons with the armchair boundary
conditions.

\subsection{Armchair ribbons}
\label{subsec:armachair_solutions}

We seek a solution in the form $\Phi=e^{ik_y y}(\phi_{A}(x),\phi_C(x),\phi_B(x);\phi_{B}'(x),\phi_C'(x),\phi_A'(x))$. The wave
functions in the $K$ valley satisfy the equations
\begin{align}\label{eq:armchair_eigeneq}
	\left(\begin{array}{ccc}
	0 & -i\p_x-ik_y & 0\\
	-i\p_x+ik_y & 0 & -i\p_x-ik_y\\
	0 & -i\p_x+ik_y & 0
	\end{array}\right)\left(\begin{array}{c}
	\phi_A(x)\\
	\phi_C(x)\\
	\phi_B(x)
	\end{array}\right)=\tilde{\epsilon}\left(\begin{array}{c}
	\phi_A(x)\\
	\phi_C(x)\\
	\phi_B(x)
	\end{array}\right).
\end{align}
The wave function in the $K^{\prime}$ valley satisfies the same equation with the replacement $\tilde{\epsilon} \to -\tilde{\epsilon}$ and the inverse order
of components. The armchair boundary conditions are given in Eq.\eqref{armchair-BC}.
\begin{figure}[h]
	\centering
	\includegraphics[scale=0.45]{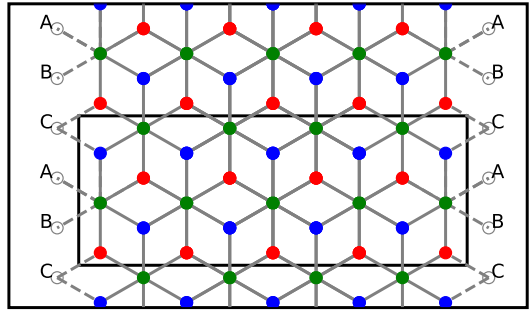}
	\caption{Ribbon with the armchair boundary conditions. The unit cell for which tight-binding calculations are performed is shown as a black
	rectangle.}
	\label{fig6}
\end{figure}
For $\tilde{\epsilon}\neq0$, we can express the $\phi_A$ and $\phi_B$ components through $\phi_C$ by using Eq.\eqref{eq:armchair_eigeneq} in
both valleys. Then the second equation in system (\ref{eq:armchair_eigeneq}) gives the equation for $\phi_C$ (the same equation is valid
for $\phi_{C}^{\prime}$ too)
\begin{align}
	\frac{\tilde{\epsilon}^2}{2}\phi_C=-(\p_x^2-k_y^2)\phi_C.
\end{align}
Its general solution is given by Eq.\eqref{zigzag:phi_C-sol}. The boundary conditions lead to the following system of equations for constants
$A,\, B,\,A',\,B'$ with $z=\sqrt{\tilde{\epsilon}^2/2-k_{y}^{2}}$:
\begin{align}
	&A+B=A'+B',\quad Ae^{izL}+Be^{-izL}=A'e^{izL+i\Delta KL}+B'e^{-izL+i\Delta KL};\\
	&A-B=-A'+B',\quad
	Ae^{izL}-Be^{-izL}=-A'e^{izL+i\Delta KL}+B'e^{-izL+i\Delta KL}.
\end{align}
This system of linear homogeneous equations has nontrivial solutions when
\begin{align}
	8 e^{i\Delta KL}(\cos(\Delta KL)-\cos(2Lz))=0.
\end{align}
The solutions to the above equation are $2zL=\pm \Delta KL+2\pi n$. Note that by definition $z\geq 0$, which gives limits on $n$.
Combining the "$+$" and "$-$" solutions gives the energy spectrum
\begin{align}\label{eq:armachair-spectrum}
	\frac{1}{2}\tilde{\epsilon}^2=k_y^2+\left(\frac{\pi n}{L}-\frac{\Delta K}{2}\right)^2
\end{align}
with integer $n=0,\pm 1,\pm 2, \dots$ and the wave functions
\begin{eqnarray}
	\Psi_n^{\vec{K}}(k_y,x)&=&\frac{1}{\tilde{\epsilon}_n\sqrt{2L}}\left(\begin{array}{c}
	-\frac{\Delta K}{2}+\frac{\pi n}{L}-ik_y\\
	\tilde{\epsilon}_n\\
	-\frac{\Delta K}{2}+\frac{\pi n}{L}+ik_y
	\end{array}\right)e^{i(-\frac{\Delta K}{2}+\frac{\pi n}{L})x+ik_y y},\,\,\nonumber\\
\Psi_n^{\vec{K}^{\prime}}(k_y,x)&=&\frac{1}{\tilde{\epsilon}_n\sqrt{2L}}\left(\begin{array}{c}
	-\frac{\Delta K}{2}+\frac{\pi n}{L}+ik_y\\
	\tilde{\epsilon}_n\\
	-\frac{\Delta K}{2}+\frac{\pi n}{L}-ik_y
	\end{array}\right)e^{-i(-\frac{\Delta K}{2}+\frac{\pi n}{L})x+ik_y y}.
\end{eqnarray}

The solutions are plain waves like in graphene [\onlinecite{Brey}].

The length $L$ is defined as $L=\frac{\sqrt{3}}{2}(\tilde{N}+1)a$ for a strip with $\tilde{N}$ atomic rows. For $L$ such that
$\Delta K L=2\pi N$ with integer $N$, the gap in spectrum Eq.(\ref{eq:armachair-spectrum}) vanishes when $\tilde{N}=3N-1$.
In this case, the spectrum contains two gapless (semi-metallic) modes with the linear dispersion
$\tilde{\epsilon}=\pm \sqrt{2}k_y$. The other energy levels have band gaps $\sim 1/L$ and are doubly degenerate. Ribbons with
$\tilde{N} \ne 3N-1$ have nondegenerate states and do not possess zero energy modes, hence these ribbons are band insulators.
In general, for armchair ribbons, we have the results similar to graphene [\onlinecite{Brey}] except the existence of
the zero-energy flat band inherent to the dice lattice model.

\subsection{Zero energy}
\label{subsec:armchair-epsilon-0}

For the zero energy $\epsilon=0$, we have again only one equation for the two components in each valley
\begin{align}
	&(-\p_x+k_y)\phi_A(x)+(-\p_x-k_y)\phi_B(x)=0,\quad (-\p_x+k_y)\phi_B'(x)+(-\p_x-k_y)\phi_A'(x)=0
\label{equation-1}
\end{align}
with the boundary conditions \eqref{armchair-BC} for $\phi_A, \phi_A'$ and $\phi_B,\phi_B'$ functions. We seek the solution in the
form
\begin{align}
	&\phi_A(x)=A_1 e^{izx} + A_2 e^{-izx},\quad \phi_A'(x)=A_1' e^{izx} + A_2' e^{-izx},\nn
	&\phi_B(x)=B_1 e^{izx} + B_2 e^{-izx}, \quad\phi_B'(x)=B_1' e^{izx} + B_2' e^{-izx}.
\label{equations-2-3}
\end{align}
Combining Eqs.(\ref{equation-1}) and (\ref{equations-2-3}), we obtain the system
\begin{eqnarray}
&&\bigg[(-iz+ky)A_1+(-iz-k_y)B_1\bigg]e^{izx}+\bigg[(iz+k_y)A_2+(iz-k_y)B_2\bigg]e^{-izx}=0,\nonumber\\
&&\bigg[(-iz-ky)A_1'+(-iz+k_y)B_1'\bigg]e^{izx}+\bigg[(iz-k_y)A_2'+(iz+k_y)B_2'\bigg]e^{-izx}=0,
\label{armachair:AB-system}
\end{eqnarray}
which is satisfied for any $0<x<L$ when the coefficients near $e^{izx}$ and $e^{-izx}$ functions are zero.
The armchair boundary conditions at the $x=0$ and $x=L$ edges give
\begin{align}\label{armchair:cond_1}
	\left\{\begin{array}{c}
	A_1+A_2=A_1'+A_2'\\
	B_1+B_2=B_1'+B_2'
	\end{array}\right.,\quad
	 \left\{\begin{array}{c}A_1 e^{izL} + A_2 e^{-izL} = e^{i\Delta KL}[A_1' e^{izL} + A_2' e^{-izL}],\\
	B_1 e^{izL} + B_2 e^{-izL} = e^{i\Delta KL}[B_1' e^{izL} + B_2' e^{-izL}]\end{array}\right. .
\end{align}
Eqs.\eqref{armachair:AB-system} together with Eq.\eqref{armchair:cond_1} have nontrivial solutions when
\begin{align}
	32k_y^2 z^2 e^{i\Delta K L} (\cos(\Delta KL)-\cos(2zL))=0.
\end{align}
This means that the system has nontrivial solutions for $z=\pm \frac{\Delta K}{2}+\frac{\pi n}{L}$ with such integer
$n$ that $z\geq 0$. The corresponding normalized wave functions for $+$ and $-$ solutions can be combined and written as
\begin{eqnarray}
	\Psi_{0}^{\vec{K}}(z_n,k_y)&=&\frac{1}{\sqrt{2(k_y^2+z_n^2)L}}\left(\begin{array}{c}
	k_y+i z_n\\
	0\\
	k_y-i z_n
	\end{array}\right)e^{iz_nx+ik_y y},\nonumber\\	
\Psi_{0}^{\vec{K}^{\prime}}(z_n,k_y)&=&\frac{1}{\sqrt{2(k_y^2+z_n^2)L}}\left(\begin{array}{c}
	k_y-i z_n\\
	0\\
	k_y+iz_n
	\end{array}\right)e^{-i z_n x+ik_y y},
\end{eqnarray}
where we used short-hand notation $z_n=-\frac{\Delta K}{2}+\frac{\pi n}{L}$ with $n=0,\pm 1,\dots$.
Hence the flat band with zero energy has infinite degeneracy parameterized by quantum numbers $k_y$ and $n$.

\subsection{Numerical results}
\label{sec:armchair-numeric}

For ribbons with the armchair edges, we compare the energy spectrum \eqref{eq:armachair-spectrum} with the results of tight-binding
calculations in Fig.\ref{fig7}, where  $L=\frac{\sqrt{3}}{2}(\tilde{N}+1)a$ with $\tilde{N}$ atomic rows. The theoretical results
are plotted as red dashed lines only in the upper energy half-plane for the clarity of presentation. The corresponding curves in
the lower half-plane trivially follow from the particle-hole symmetry. As was mentioned before, only ribbons with $\tilde{N}=3N-1$
demonstrate metallic type of spectrum, which contain gapless states with linear dispersion (see the right panel in Fig.\ref{fig7}).
The gapped (semiconducting) states are arranged in pairs with very small gaps between them for wide ribbons, while the continuum model
predicts double degeneracy of these states.  The spectrum for ribbons with a number of atomic rows different from $3N-1$ fits very well
the spectrum of continuum model (see left panel in Fig.\ref{fig7}). Similar situation is valid for graphene [\onlinecite{Brey}] where,
of course, the zero-energy flat band is absent.

\begin{figure}[h!]
	\centering
	\includegraphics[scale=0.4]{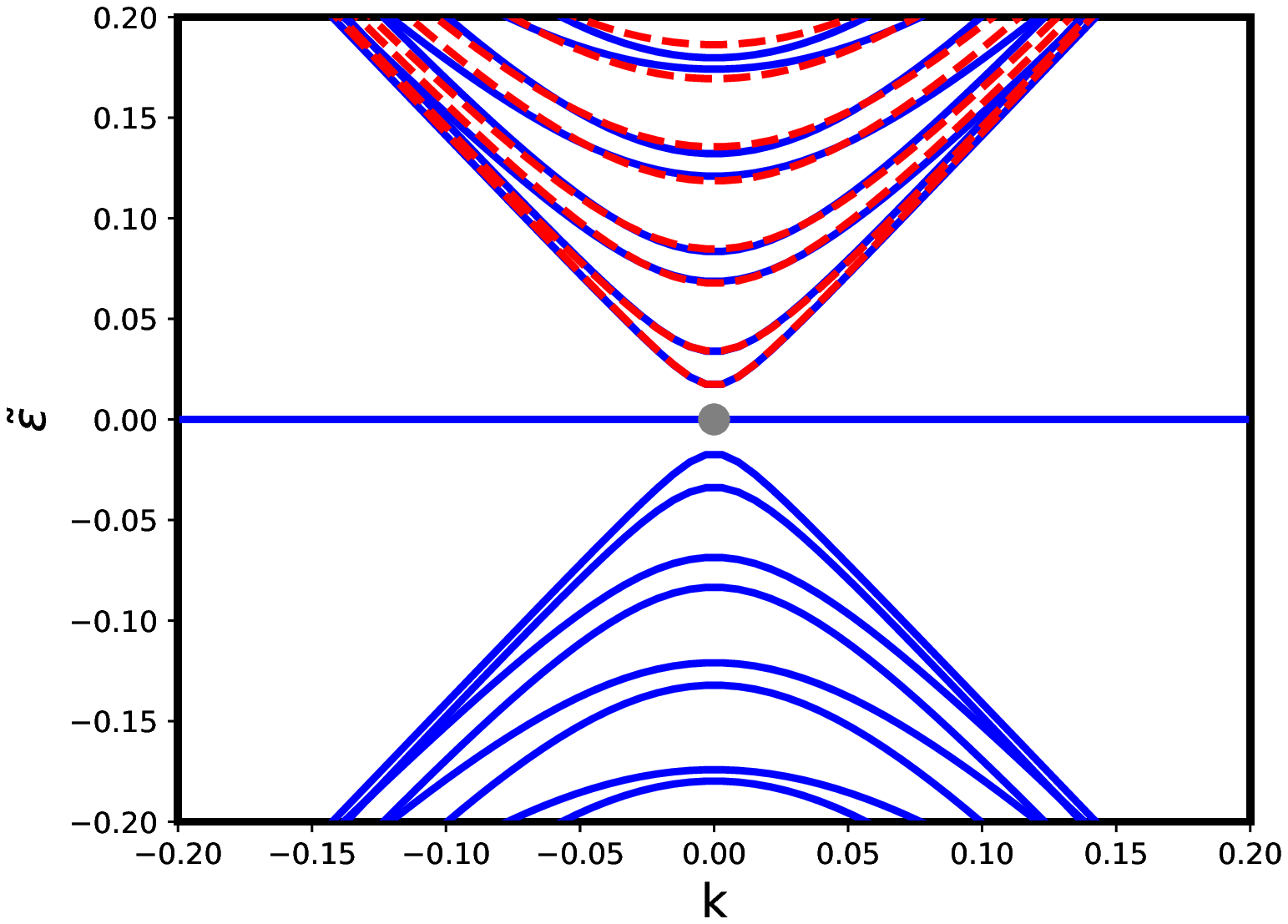}
	\includegraphics[scale=0.4]{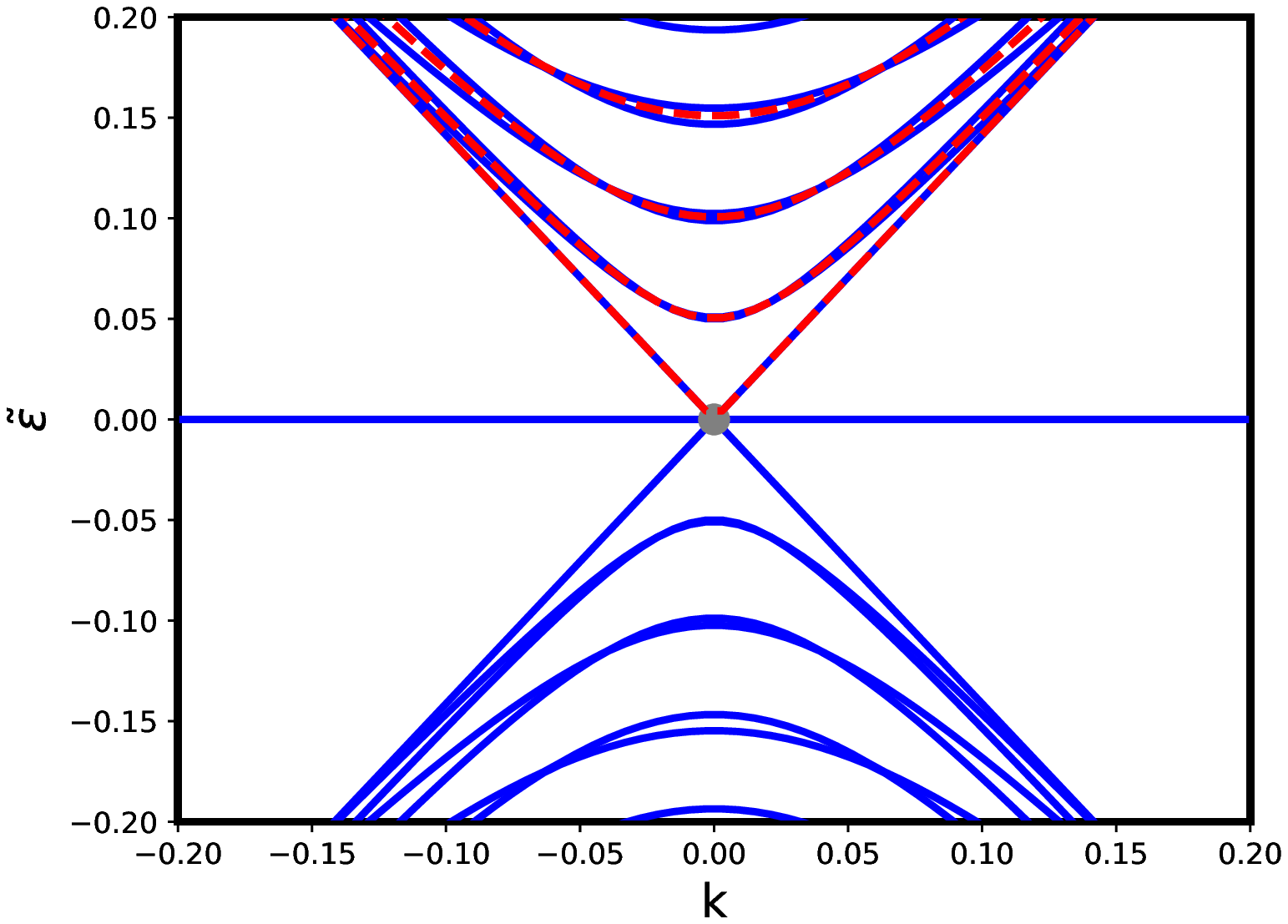}
	\caption{The numerical results (blue solid curves) and the energy dispersion given by Eq.\eqref{eq:armachair-spectrum} (red
	dashed curves in the upper energy half-plane) for a ribbon with the armchair edges. The left panel demonstrates insulating spectrum
    for a strip with 100 atomic rows. The right panel shows semi-metallic spectrum for a strip with 101 atomic rows. The gray point
    denotes the K-point.}
	\label{fig7}
\end{figure}

\section{Summary}
\label{sec:summary}

We studied the possible lattice terminations in the dice model and determined the corresponding boundary conditions. We found that there
are four possible non-equivalent zigzag terminations, but they produce in the low energy continuum model only two different types of
low-energy boundary conditions. As to the armchair boundary condition, it is unique. All these types of boundary conditions preserve
the charge conjugation and time reversal symmetries. We found the most general $6\times6$ matrix $M$ which determines boundary conditions
for the wave function of the Dirac-like equation for pseudospin-1 fermions in continuum model which extends the form of analogous matrix
for graphene [\onlinecite{Akhmerov}].

We determined the energy spectrum of ribbons with the zigzag and armchair edges. We found that in some cases the presence of boundaries
opens an energy gap between the zero-energy band and the first discrete level and leads to an insulating behavior of the system. While
the energy levels for a ribbon with armchair boundary conditions show the same features as in graphene [\onlinecite{Akhmerov, Brey}]
(except, of course, the zero-energy flat band absent in graphene), the results for ribbons with the zigzag boundary conditions are quite
different. In particular, in the dice lattice ribbons there are no propagating edge states localized at a zigzag boundary. On the other
hand, there are ribbons with specific terminations which contain modes of metallic type in a bulk.

Our numerical calculations in the tight-binding model for wide ribbons excellently confirm the analytic results obtained in the low energy
continuum model. Moreover, the qualitative structure of the energy levels in both models agrees also, although there some quantitative
differences at wave vectors far from the $K$ and $K^{\prime}$ points. We found that the zero-energy flat band in the dice lattice model
is very robust. Our calculations show that it exists for both zigzag and armchair dice lattice terminations. The boundary conditions affect
only the degeneracy of this band which is quantified by the wave vector along the termination side and an integer quantum number $n$. It
was already known [\onlinecite{Raoux,Vidal}] that the zero-energy flat band survives even in the presence of an external magnetic field which
breaks both the time reversal and charge conjugation symmetries. This clearly differs from the case of graphene in a magnetic field, where the
flat Landau levels in infinite system are deformed by the finite size of the system (see, for example, Refs.[\onlinecite{FNT2008,Gusynin2009}]).

It would be interesting to study the effects of external electric and magnetic fields in the dice model. Some of them for infinite dice lattice
in a magnetic field are already described in the literature [\onlinecite{Raoux,Malcolm}] but not for ribbons. Other effects, like the Schwinger
particle-hole pair creation [\onlinecite{Allor}] or Klein tunneling [\onlinecite{Klein-tunneling,GeimRMP2009}] in electric field wait for their
study for pseudospin-1 fermions. Also, the electronic states of pseudospin-1 fermions in the field of charged impurities are of
considerable interest (for a similar study in graphene, see, for example, review [\onlinecite{Gorbar2018FNT}]).

\begin{acknowledgments}
The work of E.V.G. and V.P.G. is partially supported by the National Academy of Sciences of Ukraine (project 0116U003191) and by its Program of
Fundamental Research of the Department of Physics and Astronomy (project No. 0117U000240). V.P.G. acknowledges the support of the RISE Project
CoExAN GA644076.
\end{acknowledgments}

\appendix

\section{Derivation of general boundary condition}
\label{sec:M_parametric}

It is convenient to represent $M$ in the basis
\begin{align}\label{eq:M_general_form}
M=\sum\limits_{\mu=0}^{3}\sum\limits_{\nu=0}^{8}(\tau_{\mu} \otimes\lambda_{\nu}) c_{\mu\nu},
\end{align}
where the coefficients $c_{\mu\nu}$ are real because matrix $M$ is Hermitian.

\subsection{General form of matrix M}
\label{subsec:commutator}

By using the property $(A\otimes B)(C\otimes D)=(AC)\otimes(BD)$, we easily find that vanishing of the anticommutator of $M$ with the normal
component of the electric current at a boundary $\{M,\mathbf{n}\mathbf{J}\}=0$ gives
\begin{align}
\sum\limits_{\mu=0}^{3}\sum\limits_{\nu=0}^{8}
\bigg[(\tau_{\mu}\tau_3) \otimes(\lambda_{\nu}(\vec{S}\vec{n}))+(\tau_3\tau_{\mu})\otimes((\vec{S}\vec{n})\lambda_{\nu})\bigg] c_{\mu\nu}=0.
\end{align}
Since $\tau_3$ commutes with the $\tau_0$ and $\tau_3$ matrices and anticommutes with $\tau_1$ and $\tau_2$, we obtain the following equations
for $c_{\mu\nu}$:
\begin{align}
\label{g_c_eq:c03}
&\sum\limits_{\mu=0}^{8}\bigg\{\lambda_{\mu},\,\, (\vec{S}\vec{n})\bigg\}c_{(0,3)\mu}=0,\\
\label{g_c_eq:c12}	
&\sum\limits_{\nu=0}^{8}\bigg[\lambda_{\nu},\,\, (\vec{S}\vec{n})\bigg]c_{(1,2)\nu}=0,
\end{align}
or explicitly in terms of the Gell-Mann matrices,
\begin{align}
&\sum\limits_{\mu=0}^{8}\bigg\{\lambda_{\mu},\,\, (\lambda_1+\lambda_6)n_x+(\lambda_2+\lambda_7)n_y\bigg\}c_{(0,3)\mu}=0,\\
&\sum\limits_{\nu=0}^{8}\bigg[\lambda_{\nu},\,\, (\lambda_1+\lambda_6)n_x+(\lambda_2+\lambda_7)n_y\bigg]c_{(1,2)\nu}=0.
\end{align}
Calculating the anticommutator in the first equation and the commutator in the second, we obtain two matrix equations. Further, setting the
coefficients at different Gell-Mann matrices to zero, we find the following system of equations for the coefficients of matrix $M$:
\begin{align}
\label{system:general_f-g_3}	
&n_y f_2=-n_x f_1,\,\,f_4=\frac{2}{3}(n_{x}^{2}-n_{y}^{2})(-3f_0+f_3),\,\,
f_5=\frac{4}{3}n_x n_y(-3f_0+f_3),\,\, f_6=-f_1,\,\,n_y f_7=n_x f_1,\,\,f_8=-\frac{1}{\sqrt{3}}f_3,\nn
&n_x g_2=n_y g_1,\,\,\,g_4=2(n_y^2-n_x^2)g_3,\,\,\,g_5=-4n_x n_y g_3,\,\,\,g_6=g_1,\,\,\,
n_x g_7=n_y g_1,\,\,\,g_8=-\frac{1}{\sqrt{3}}g_3,
\end{align}
where we used the notation $c_{(0,3),\mu}=f_{\mu}$ and $c_{(1,2),\nu}=g_{\nu}$. Thus, we have 3-parametric family of $f_{\mu}$ and $g_{\nu}$
which defines a 12-parametric family of $M$-matrices. The condition $M^2=1$ further reduces the number of parameters leaving only
six independent ones.

\subsection{Symmetry restrictions}

The Hamiltonian of the dice model is invariant with respect to the time reversal $T$ and charge conjugation $C$ transformations.
The operator $T$ has the form $\hat{T}=\tau_1\otimes F\hat{K}$ and $\hat{K}$ is the operator of complex conjugation.
The relation
\begin{align}
\hat{T}H(\vec{k})\hat{T}^{-1}=H(-\vec{k})
\end{align}
implies the two following equations for $F$:
\begin{align}
FS_x F^{-1}=S_x,\quad FS_y F^{-1}=-S_y,
\end{align}
whose solution with $F^{\dagger}F=1$ and up to an arbitrary phase factor is
\begin{align}
F=\left(\begin{array}{ccc}
0 & 0& 1\\
0& 1 & 0\\
1 &0 & 0
\end{array}\right).
\label{F-matrix}
\end{align}
The time reversal operator $\hat{T}$ satisfies $\hat{T}^2=1$. Note that in the presence of a real spin degree of freedom the operator
$\hat{T}$ should be replaced by  the operator $\hat{\mathcal{T}}=i\sigma_{2}\otimes\hat{T}$ which satisfies the standard condition
$\hat{\mathcal{T}}^2=-1$ with the matrix $\sigma_2$ acting in real spin space.

Clearly, $\hat{S}$ is a symmetry if $M\hat{S}=\hat{S}M$. Using the general form of matrix $M$ given by Eq.\eqref{eq:M_general_form},
we find that the time reversal symmetry  leads to
\begin{align}
M (\tau_1\otimes F)-(\tau_1\otimes F) M^{*}=0
\end{align}
that gives for real $c_{\mu\nu}$
\begin{align}
\sum_{\mu=0}^{3}\sum_{\nu=0}^{8}\bigg((\tau_{\mu}\tau_1)\otimes(\lambda_{\nu} F)-(\tau_1\tau^{*}_{\mu})\otimes(F\lambda^{*}_{\nu})\bigg)c_{\mu,\nu}=0.
\end{align}
The above equation implies
\begin{align}
\label{eq_condition:t-reverse_1}
\sum\limits_{\nu=0}^{8}\,(\lambda_{\nu} F-F\lambda_{\nu}^*)\,c_{\mu,\nu}=0\,\,\Rightarrow\,\,c_{\mu,1}
=c_{\mu,6},\,\,c_{\mu,2}=c_{\mu,7},\,\,c_{\mu,8}=-\frac{1}{\sqrt{3}}c_{\mu,3},
\end{align}
for $\mu=0,1,2$, and
\begin{align}
\label{eq_condition:t-reverse_2}	
\sum\limits_{\nu=0}^{8}\,(\lambda_{\nu} F+F\lambda_{\nu}^*)\,c_{3,\nu}=0\,\,\Rightarrow\,\,c_{3,0}
=c_{3,4}=c_{3,5}=0,\,\,c_{3,1}=-c_{3,6},\,\,c_{3,2}=-c_{3,7},\,\,c_{3,8}=\sqrt{3}c_{3,3}
\end{align}
for $\mu=3$.

Combining Eq.\eqref{system:general_f-g_3} with conditions \eqref{eq_condition:t-reverse_1} and \eqref{eq_condition:t-reverse_2} we find the
set of nine survived parameters, $c_{0,0},\quad c_{0,3},\quad c_{3,1},\quad c_{(1,2),0},\quad c_{(1,2),1},\quad c_{(1,2),3}$\,,
if the condition $M^2=1$ is not taken into account. The condition $M^2=1$ gives one more constraint and leaves free only five parameters.

The operator of the charge conjugation does not interchange spinors from different $K$ valleys, therefore, it is defined by the equation
\begin{align}
\hat{C} H(\vec{k})\hat{C}^{-1}=-H(\vec{k}),
\end{align}
whose solution is
\begin{align}
\hat{C}=\tau_0\otimes\left(\begin{array}{ccc}
1& 0& 0\\
0& -1& 0\\
0 & 0& 1
\end{array}\right)=\tau_0\otimes\left(\frac{1}{3}\lambda_0+\lambda_3-\frac{1}{\sqrt{3}}\lambda_8\right).
\label{C-operator}
\end{align}
Using the general form of matrix $M$ given by Eq.\eqref{eq:M_general_form}, we find that the charge conjugation symmetry leads to the following
equation for $M$:
\begin{align}
M\hat{S}_{C}-\hat{S}_C M=0\quad\Rightarrow\quad \sum_{\mu=0}^{3}\sum_{\nu=0}^{8}\tau_{\mu}\otimes \left[\lambda_{\nu},
\left(\frac{1}{3}\lambda_0+\lambda_3-\frac{1}{\sqrt{3}}\lambda_8\right)\right]c_{\mu,\nu}=0,
\end{align}
which for every $\mu=0,\,1,\,2,\,3$ gives the following restrictions on parameters:
\begin{align}
\label{eq_condition:elec-hole_symmetry}
c_{\mu,1}=c_{\mu,2}=c_{\mu,6}=c_{\mu,7}=0.
\end{align}
According to Eq.\eqref{system:general_f-g_3}, there remain independent only eight parameters, which can be chosen as $c_{0,0}, \, c_{3,0},
\, c_{0,3}, \, c_{3,3}, \, c_{1,0}, \, c_{2,0}, \, c_{1,3}, \, c_{2,3}$ (without taking into account the constraints $M^2=1$).

The general form of matrix M, which preserves T- and C-symmetries is:
\begin{eqnarray}\label{M_s_t-general}
M_{S_T,\,S_C}&=&\tau_0\otimes\bigg(c_{0,0}\lambda_0+c_{0,3}\lambda_3+\frac{2}{3}(n_x^2-n_y^2)(-3c_{0,0}+c_{0,3})\lambda_4+\frac{4}{3}n_xn_y(-3c_{0,0}
+c_{0,3})\lambda_5-\frac{1}{\sqrt{3}}c_{0,3}\lambda_8 \bigg)+\nn \nonumber\\
&+&\sum\limits_{i=1}^{2}\tau_i\otimes\bigg(\lambda_0 c_{i,0}+(\lambda_3+2(n_y^2-n_x^2)\lambda_4-4n_x n_y \lambda_5-\frac{1}{\sqrt{3}}\lambda_8)c_{i,3}\bigg).
\end{eqnarray}
Note that this expression contains six independent parameters. The condition $M^2=1$ further restricts the number of free parameters giving
several families of solutions with a maximal subset having two parameters.

\section{Boundary conditions from zero boundary current}
\label{appendix:zero-current}

We start with the exact formula for the matrix element of current \eqref{eq:current}
\begin{eqnarray}
	\frac{1}{v_F}\left\langle\Psi\right|\vec{n}_B\vec{J}\left|\Psi\right\rangle &=&\bigg[\Psi_{A}^{*} n_{-}\Psi_C +\Psi_{C}^{*}(n_+\Psi_A+n_{-}\Psi_B)+\Psi_{B}^{*}n_{+}\Psi_{C}\bigg]\nonumber\\
&-&\bigg[\Psi_{B'}^{*} n_{-}\Psi_{C'} +\Psi_{C'}^{*}(n_+\Psi_{B'}+n_{-}\Psi_{A'})+\Psi_{A'}^{*}n_{+}\Psi_{C'}\bigg],
\end{eqnarray}
where $n_{\pm}=n_x\pm in_y$. We begin with the zigzag boundary conditions $n_x=0,\,\,n_y=1$.

1. {\it Zigzag boundary conditions.}
For the zigzag boundary conditions, it is sufficient to analyze only one valley. The matrix element in the K valley has the form
\begin{align}
	\frac{1}{v_F}\left\langle\Psi\right|\vec{n}_B\vec{J}\left|\Psi\right\rangle\bigg|_{K}=(\Psi_{A}^{*} n_{-}+\Psi_{B}^{*}n_{+})\Psi_C +\Psi_{C}^{*}(n_+\Psi_A+n_{-}\Psi_B)=0.
	\label{normal-current-zigzag-1}
\end{align}
Obviously, the above equation has two possible solutions for $n_y=\pm 1$:
\begin{align}
\Psi_A-\Psi_B=0 \quad and \quad \Psi_C=0.
\label{normal-current-zigzag-2}
\end{align}
This means that the \textbf{AB} boundary condition can be written as $\Psi_A-\Psi_B=0$.

For the \textbf{CA} or \textbf{BC} boundary conditions, we automatically have missing C-atoms. Then, the simplest boundary condition is
$\Psi_C=0$. There are also some corrections from missing A- or B-atoms, but we can neglect them in our analysis, because these corrections
are important only for upper levels, where the linearized Hamiltonian cannot be applied.

2. {\it Armchair boundary conditions.}
Here we need to combine both valleys. Imposing the armchair boundary conditions on the $x$ sides, ($n_{B}=(\pm 1,\, 0)$), we find for
the matrix element of the current
\begin{align}
\frac{1}{v_F}\la\Psi\right|\vec{n}_B\vec{J}\left|\Psi\ra =\bigg[(\Psi_{A}^{*}+\Psi_{B}^{*}) \Psi_C +\Psi_{C}^{*}(\Psi_A+\Psi_B)\bigg]
-\bigg[(\Psi_{A'}^{*}+\Psi_{B'}^{*}) \Psi_{C'} +\Psi_{C'}^{*}(\Psi_{A'}+\Psi_{B'})\bigg]=0.
\end{align}
Therefore, the possible types of conditions are
\begin{align}
	&\Psi_{\mu}=e^{i\alpha}\Psi_{\mu'},\quad \Psi_{\mu}=e^{i\alpha}\Psi_{\mu'}^{*} \quad\mu=A,B,C
\end{align}
with real phase $\alpha$, which is equal to all three functions (this phase cancels out due to complex conjugation in
the products).

\newpage

\end{document}